# Verifying Quantum Circuits with
# Level-Synchronized Tree Automata (Technical Report)

**PAROSH AZIZ ABDULLA**, Uppsala University, Sweden
**YO-GA CHEN**, Academia Sinica, Taiwan
**YU-FANG CHEN**, Academia Sinica, Taiwan
**LUKÁŠ HOLÍK**, Brno University of Technology, Czech Republic
**ONDŘEJ LENGÁL**, Brno University of Technology, Czech Republic
**JYUN-AO LIN**, National Taipei University of Technology, Taiwan
**FANG-YI LO**, Academia Sinica, Taiwan
**WEI-LUN TSAI**, Academia Sinica, Taiwan and National Taiwan University, Taiwan

We present a new method for the verification of quantum circuits based on a novel symbolic representation of sets of quantum states using *level-synchronized tree automata* (LSTAs). LSTAs extend classical tree automata by labeling each transition with a set of *choices*, which are then used to synchronize subtrees of an accepted tree. Compared to the traditional tree automata, LSTAs have an incomparable expressive power while maintaining important properties, such as closure under union and intersection, and decidable language emptiness and inclusion. We have developed an efficient and fully automated symbolic verification algorithm for quantum circuits based on LSTAs. The complexity of supported gate operations is at most quadratic, dramatically improving the exponential worst-case complexity of an earlier tree automata-based approach. Furthermore, we show that LSTAs are a promising model for *parameterized verification*, i.e., verifying the correctness of families of circuits with the same structure for any number of qubits involved, which principally lies beyond the capabilities of previous automated approaches. We implemented this method as a C++ tool and compared it with three symbolic quantum circuit verifiers and two simulators on several benchmark examples. The results show that our approach can solve problems with sizes orders of magnitude larger than the state of the art.

## 1 INTRODUCTION

With the recent progress in quantum hardware and the push to achieve *quantum supremacy* (cf., e.g., [6]), the vision of quantum computing, an ability to solve practical conventionally unsolvable problems, is slowly getting real. Systems and languages for programming quantum computers are

---

Authors' addresses: Parosh Aziz Abdulla, Uppsala University, Department of Information Technology, Sweden, parosh@it.uu.se; Yo-Ga Chen, Academia Sinica, Institute of Information Science, Taiwan, googagagaga@gmail.com; Yu-Fang Chen, Academia Sinica, Institute of Information Science, Taiwan, yfc@iis.sinica.edu.tw; Lukáš Holík, Faculty of Information Technology, Brno University of Technology, Czech Republic, holik@fit.vut.cz; Ondřej Lengál, Faculty of Information Technology, Brno University of Technology, Czech Republic, lengal@fit.vut.cz; Jyun-Ao Lin, iFIRST & CSIE, National Taipei University of Technology, Taiwan, jyalin@gmail.com; Fang-Yi Lo, Academia Sinica, Institute of Information Science, Taiwan, lofangyi@gmail.com; Wei-Lun Tsai, Academia Sinica, Institute of Information Science, Taiwan, alan23273850@gmail.com and National Taiwan University, Graduate Institute of Electronics Engineering, Taiwan.

being intensively developed [3, 27, 57], together with efficient quantum algorithms for solving real-world problems such as machine learning [10, 20], optimization [42],

This progress drives a demand for tools for reasoning about quantum programs. Writing quantum programs is indeed immensely challenging due to their probabilistic nature and the exponential size of the computational space. Errors are easily made and difficult to find. Tools for reasoning about quantum programs would ideally have the following properties: (1) *Flexibility:* allows flexible specification of properties of interest, (2) *Diagnostics:* provides precise bug diagnostics, (3) *Automation:* operates automatically, and (4) *Scalability:* scales efficiently to verify useful programs.

*Symbolic verification* [2, 12, 35, 36, 43] is one of the most successful techniques that satisfies the above criteria for conventional programs. However, there has been minimal progress in the area of symbolic verification for quantum circuits. This paper contributes towards filling this gap by adapting automata-based symbolic verification [2] to quantum circuits. More specifically, we think of the symbolic verification problem in terms of Hoare triples {P} C {Q}, where P and Q are sets of quantum states representing the *pre-condition* and the *post-condition*, and $C$ is a quantum circuit. We aim to use symbolic verification to check the validity of the triple, ensuring that all executions of $C$ from states in P result in states within Q.

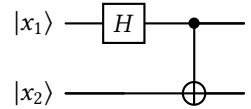

Fig. 1. The Bell state circuit. ● denotes the control qubit and ⊕ denotes the target qubit on which X is applied.

*Example 1.* As an example of a verification problem, let $C$ be the circuit creating Bell states in Fig. 1. The circuit begins by applying the H gate to the first qubit, followed by a controlled-X gate with the first qubit as the control and the second qubit as the target. The circuit converts 2-qubit computational basis states to Bell states (maximally entangled states). Its correct implementation satisfies the Hoare triple with the pre-condition P = {|00⟩, |01⟩, |10⟩, |11⟩} (computational basis states are denoted using the Dirac notation |$x_1 x_2$⟩) and the post-condition Q = $\{\frac{1}{\sqrt{2}}(|00\rangle \pm |11\rangle), \frac{1}{\sqrt{2}}(|01\rangle \pm |10\rangle) \mid \pm \in \{+, -\}\}$ (Bell states).    □

Our quantum circuit symbolic verification framework comprises three major components: (1) a *succinct symbolic representation of sets of quantum states* allowing efficient manipulation and facilitating flexible specification of desired pre- and post-conditions, (2) algorithms for *symbolic execution of individual quantum gates*, which compute the symbolic representation of output states from input states and explore the reachable state space after circuit execution, (3) *entailment test* on the symbolic representation to verify that all reachable states conform to the specified post-condition or report a witness when the specification is violated. Together, these components enable *flexible*, *diagnostic*, *automated*, and *scalable* verification of quantum circuits.

*Succinct symbolic representation.* The central part of our framework is a novel symbolic representation of sets of quantum states, the *level-synchronized tree automata (LSTAs)*. The concept evolved from decision diagrams that have been used to succinctly represent one quantum state, e.g., in [41, 48, 50, 54]. They are based on viewing a quantum state as a binary tree, where each branch (a path from a root to a leaf) represents a *computational basis state*, such as |10⟩ or |00⟩ for a two-qubit circuit. The tree is *perfect*, i.e., the length of every branch is the same and equals

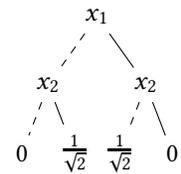

Fig. 2. $\frac{1}{\sqrt{2}}(|01\rangle + |10\rangle)$.

the number of qubits in the circuit. The leaves represent the *complex probability amplitudes*[1] of the state. An example in Fig. 2 shows the tree representing a state with two qubits where the basis states $|01\rangle$ and $|10\rangle$ have the probability amplitude $\frac{1}{\sqrt{2}}$ and the others have the amplitude 0. In the figure, dashed edges denote the 0 value and solid edges denote the 1 value of the variable that is in the source node of the edge, so the $|01\rangle$ state is encoded by first taking the dashed edge (from the node labelled by $x_1$) and then the solid one (from the left-most $x_2$ node).

---

[1]Amplitude is a generalization of the concept of "probability." The square of the absolute value of a complex amplitude represents the corresponding probability. The use of complex numbers allows for the expression of "negative probabilities" (obtained after squaring) that are canceled out due to interference.

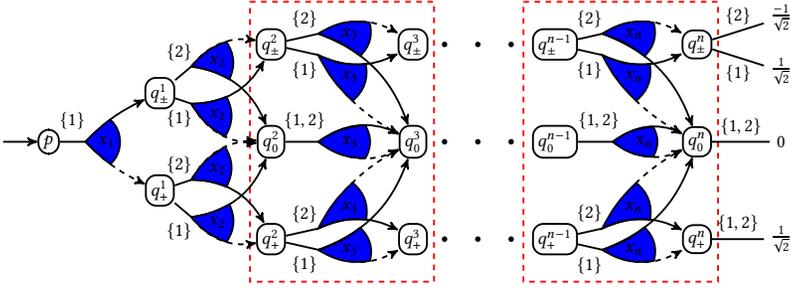

Fig. 4. An LSTA representing the set of quantum states of the form $\frac{1}{\sqrt{2}}(|0b_2b_3\dots b_n\rangle \pm |1\bar{b}_2\bar{b}_3\dots\bar{b}_n\rangle)$.

LSTAs enrich decision diagrams with three important features. First, they allow *disjunctive branching*, which increases their compactness, enabling different quantum states to share common structures. This has also been observed and used in the work of [18], resulting in an exponential space saving compared with storing quantum states as a set of BDDs. Second, they allow *cycles*, which enables representing unboundedly many quantum states and opens a path towards verification of quantum circuits with a parameterized number of qubits. The addition of cycles and disjunctive branching results in a class of tree automata. On top of that, LSTAs are equipped with a novel mechanism of tree *level synchronization*, which yet again dramatically increases succinctness (up to exponentially) and simplifies the symbolic execution of quantum gates.

*Example 2 (An LSTA for a simple 2-qubit circuit.).* We intuitively explain the features of LSTAs on an encoding of the post-condition from Example 1. The set of Bell states (post-condition) is generated by the LSTA in Fig. 3. The LSTA generates trees representing quantum states from the root downward, starting at the root state $p$, and proceeding iteratively by picking a transition to generate children states, until reaching the leaves. For example, the tree from Fig. 2 can be generated by first picking the transition $(p) \xrightarrow{\{1\}} \blacktriangleright (q_+)(q_\pm)$, then the two transitions, $(q_+) \xrightarrow{\{2\}} \blacktriangleright (r_0)(r_+)$, $(q_\pm) \xrightarrow{\{2\}} \blacktriangleright (r_\pm)(r_0)$, and ending with the leaf transitions $(r_+) \xrightarrow{\{1,2\}} (\frac{1}{\sqrt{2}})$, $(r_0) \xrightarrow{\{1,2\}} (0)$, $(r_\pm) \xrightarrow{\{1\}} (\frac{1}{\sqrt{2}})$. Like the traditional *tree automata* (TAs) model [22], LSTAs allow disjunctive branches and use them to express multiple states with a shared structure. Here, the states $q_+, q_\pm$, and $r_\pm$ have two disjunctive transitions. Their combination would generate 8 different trees. Not all of these combinations are, however, intended. LSTAs enrich the traditional TA model by labeling the transitions with sets of *choices* (the $\{1\}$, $\{2\}$, and $\{1,2\}$ in this example). The choices play an essential role in restricting the set of generated trees to the intended ones. Namely, at every tree level, the used transitions must agree on a common choice, otherwise the tree will be rejected. Particularly in the second level of the tree, the transitions labeled $\{1\}$ can be taken together, as they agree on 1, or transitions labeled {2} can be taken together, as they agree on 2. A combination of a transition labeled {1} and a transition labeled {2} is not admissible as their sets of choices are disjoint. Including also the admissible choices at the third level, the LSTA generates exactly the 4 Bell states (using the 9 transitions in the figure). □

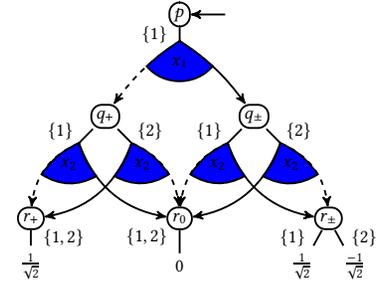

Fig. 3. The LSTA representing Bell states

*Example 3 (Succinctness of LSTAs in a larger $n$-qubit circuit).* The succinctness of LSTAs and the role of the level synchronization is visible when the previous example is generalized to $n$ qubits, where LSTAs can represent the $2^n$ output quantum states with *a linear number of transitions*. Indeed, the 2-qubit circuit can be generalized to $n$-qubit circuits, generating the so-called GHZ states[2] [28] $Q = \{\frac{1}{\sqrt{2}}(|0b_2b_3\dots b_n\rangle \pm |1\bar{b}_2\bar{b}_3\dots\bar{b}_n\rangle) \mid b_2b_3\dots b_n \in \mathbb{B}^{n-1} \wedge \pm \in \{+,-\}\}$, where $\bar{b}$ denotes the complement of $b$. We visualize a GHZ state in Fig. 5; the basis $|0b_2b_3\dots b_n\rangle$ has amplitude $\frac{1}{\sqrt{2}}$ and $|1\bar{b}_2\bar{b}_3\dots\bar{b}_n\rangle$ has amplitude $\frac{1}{\sqrt{2}}$ or $\frac{-1}{\sqrt{2}}$. An LSTA can be used to represent the post-condition $Q$ with only $5n - 1$ transitions (see Fig. 4). This results in an *exponential space saving* compared to other standard ways of storing sets of quantum states precisely, such as traditional tree automata [18], sets of BDDs [50], or sets of state vectors [37], which all need exponential space to store the $2^n$ quantum states in $Q$.

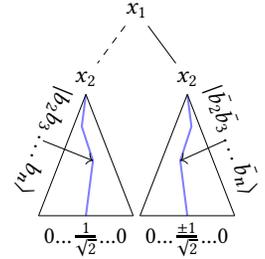

Fig. 5. GHZ states

LSTAs achieve this succinctness by combining disjunctive branching and level synchronization. Each GHZ state consists of a left-hand side 0-rooted subtree (called 0-subtree below) and a right-hand side 1-rooted subtree (called 1-subtree below), see Fig. 5. Each subtree has all leaves except one labeled 0. The two distinguished leaves are reached by two paths that are mirror images of each other, representing the 0-basis $b_2b_3\dots b_n$ and the inverted 1-basis $\bar{b}_2\bar{b}_3\dots\bar{b}_n$. Disjunctive branching does the first part of the job: it represents the set of all $2^{n-1}$ subtrees with a single distinguished leaf by a number of transitions linear to $n$. The LSTA traces the path towards the distinguished leaf by a sequence of states $q_\pm^1,\dots,q_\pm^n$ in the 0-subtree and $q_\pm^1,\dots,q_\pm^n$ in the 1-subtree (Fig. 4). When the subtrees are generated, each state on the path spawns two children. One is the next state in the sequence, the other is a state $q_0^i$ that generates a uniform tree with all leaves 0. Each state on the path can choose to continue the path to the left or to the right by choosing one of two transitions (this is the disjunctive branching).

Level synchronization is then used to ensure that the two paths towards the distinguished leaves are mirror images of each other. In every level of the tree, if the path in the 0-subtree is continuing to the left, then the path in the 1-subtree must continue to the right, and vice versa. The LSTA in Fig. 4 achieves this as follows: in the 0-subtree, the left/right transitions continuing the path are associated with the choices 1 and 2, respectively, and in the right subtree, the left/right continuing transitions are associated with the choices in the inverted manner, 2 and 1, respectively. Without this level-synchronization mechanism, each 0-subtree would need a unique root transition to connect to its corresponding 1-subtree, requiring $2^{n-1}$ root transitions (we argue in Theorem 16 that an exponential number of transitions is unavoidable). □

In fact, many quantum gates create a correspondence between subtrees of a tree (cf. Section 2.2.1 and Fig. 7)—this is a typical manifestation of quantum entanglement. Hence, level synchronization is also helpful in the general case, not only for some special circuits. We will give more examples in Section 4, where we demonstrate that LSTAs can succinctly express a wide range of correctness properties of quantum circuits, including the following verification tasks. All the involved LSTAs are of a size linear in the number of qubits.

---

[2] Often GHZ states refer to the set $\{\frac{1}{\sqrt{2}}(|0^n\rangle + |1^n\rangle) \mid n \in \mathbb{N}\}$, but here we refer to a generalized version obtained by feeding all $n$-qubit computational basis states to the GHZ generating circuit.

- We can verify the correctness of a circuit component, for example, *a multi-control Toffoli gate implemented with standard Toffoli gates (with two control inputs).*
- We can construct a template oracle circuit that reads secret strings from the input and use it to verify *oracle-based circuits.* E.g., by verifying a Bernstein-Varzirani circuit against all possible oracles, we ensure it correctly identifies the secret string with just one oracle query.
- By allowing the use of variables at tree leaves, we can verify that *a Grover iteration indeed increases the probability of finding a correct solution* for infinitely many feasible input states.
- Moreover, LSTAs can be used to specify *the equivalence of two circuits $C$ and $C'$.* We can use LSTAs to express a set of $2^n$ *linearly independent vectors* compactly using a linear (in $n$) number of transitions. If we use this LSTA as the pre-condition and also as the post-condition for a symbolic verification framework, we can check if a circuit's function corresponds to *identity*. We can then check if $C$ and $C'$ are equivalent by sequentially composing $C$ with the inverse of $C'$ and checking if the result is identity.

*Fast symbolic execution of gates.* With an LSTA-encoded precondition, the next step is to compute an LSTA encoding all states reachable from the precondition after executing a circuit. In Section 5, we have developed algorithms to execute gates symbolically, that is, to compute LSTA-represented output states from LSTA-represented input states and a single quantum gate $U$. We support a wide variety of quantum gates, including all single-qubit gates and (multi-)controlled-gates, such as the Toffoli gate. We show that all supported gates can be executed over LSTAs in a time *quadratic* in the size of the input LSTA. The reachable states can then be computed via a sequence of symbolic gate executions.

*Entailment of LSTAs and other operations.* The next step is to verify if all reachable states satisfy the postcondition. In Section 6, we present an algorithm for the entailment (i.e., language inclusion) test between LSTAs. We show that LSTA entailment is decidable with the complexity between **PSPACE** and **EXPSPACE** and can be implemented to run fast enough in practice. When the entailment test fails, the algorithm can return a tree witnessing the entailment violation. The quantum state represented by the tree can then be used to diagnose the quantum circuit and find out why verification fails. We also show that LSTAs have decidable (**PSPACE**-complete) emptiness problem, are closed under union and intersection, but not closed under complement.

*Experimental evaluation.* Our experimental results in Section 7 clearly demonstrate that LSTAs enable a highly efficient and scalable framework for automated quantum circuit verification. Implemented as an updated version of our tool, AUTOQ, our approach successfully handled multiple verification tasks, verified all specified correctness properties, and found all injected bugs. We compared the new AUTOQ with three recent symbolic quantum circuit verifiers, AUTOQ-OLD [18], CAAL [19], and SYMQV [7] and two simulators SLIQSIM [50] and SV-SIM [37] on several benchmark examples. Notably, AUTOQ significantly outperformed all other tools in these tasks.

*Towards parameterized verification of quantum circuit.* As LSTAs naturally allow cycles in their transition relation, they show a promise for *parameterized verification* of quantum circuits, checking the correctness of a *circuit template* for any (parametric) number of qubits. For instance, Fig. 6 contains an LSTA that encodes the set of states $|0^n\rangle$ for any number of qubits $n$. Here, we label transitions with $x$ to denote that it is an unspecified qubit, and its value depends on the tree level on which it is used. We extended our approach to support various types of parameterized quantum gates, including the application of CX gates to every consecutive qubit. We were able to describe the template and establish the correctness of GHZ circuits [28] and circuits performing *diagonal Hamiltonian simulation* [40] and *fermionic unitary evolution* [60], which are frequently used in quantum chemistry and material science.

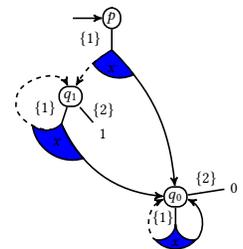

Fig. 6. $\{|0^n\rangle \mid n \geq 1\}$.

## 2  PRELIMINARIES

This section aims to provide readers with a basic understanding of quantum computing. Quantum computers are programmed through *quantum gates*, and each gate application updates the global *quantum state*. A *quantum circuit* is a sequence of quantum gate applications.

### 2.1  Quantum States

In a traditional computer system with $n$ bits, a state is represented by $n$ Boolean values. In the quantum world, such states are referred to as *computational basis states*. For example, in a system with three bits labeled $x_1$, $x_2$, and $x_3$, the computational basis state $|011\rangle$ indicates that the value of $x_1$ is 0 and the values of $x_2$ and $x_3$ are 1. In a quantum system, an $n$-qubit *quantum state* encodes the amplitude information of a superposition of all possible $n$-bit computational basis states, denoted as a formal sum $\sum_{j\in\{0,1\}^n} a_j \cdot |j\rangle$, where $a_0, a_1, \ldots, a_{2^n-1} \in \mathbb{C}$ are *complex amplitudes* satisfying the property that $\sum_{j\in\{0,1\}^n} |a_j|^2 = 1$. Intuitively, $|a_j|^2$ is the probability that when we measure the quantum state in the computational basis, we obtain the classical state $|j\rangle$; these probabilities must sum up to 1 for all computational basis states. The standard representation of a quantum state is a vector $(a_0, a_1, \ldots, a_{2^n-1})^T$ of amplitude values, where the superscript $T$ denotes *transposition*.

We represent a quantum state using a *decision tree* where each branch represents a computational basis state and the leaves hold complex amplitudes. We demonstrate in Fig. 8a an example of a decision tree encoding a quantum state $q = a \cdot |00\rangle + b \cdot |01\rangle + c \cdot |10\rangle + d \cdot |11\rangle$. This viewpoint enables us to see the definition of standard quantum gate operations as tree transformations.[3] The viewpoint also allows us to represent a set of states compactly using LSTAs. In fact, the tree view can be generalized to handle any vector of $2^n$ entries. We will show that LSTAs can compactly represent some sets of linearly independent vectors and use them for testing circuit equivalence.

### 2.2  Quantum Gates and Circuits

The two main types of quantum gates used in state-of-the-art quantum computers are *single-qubit gates* and *controlled gates*.

#### 2.2.1  General single-qubit gates.
In general, a single-qubit gate is presented as a *unitary complex matrix* U, shown below together with some common examples of this category ($\theta$ is a parameter):

$$U = \begin{pmatrix} u_1 & u_2 \\ u_3 & u_4 \end{pmatrix}, X = \begin{pmatrix} 0 & 1 \\ 1 & 0 \end{pmatrix}, Y = \begin{pmatrix} 0 & -i \\ i & 0 \end{pmatrix}, R_X(\theta) = \begin{pmatrix} \cos\frac{\theta}{2} & -i\sin\frac{\theta}{2} \\ -i\sin\frac{\theta}{2} & \cos\frac{\theta}{2} \end{pmatrix}, H = \frac{1}{\sqrt{2}}\begin{pmatrix} 1 & 1 \\ 1 & -1 \end{pmatrix},$$

$$Z = \begin{pmatrix} 1 & 0 \\ 0 & -1 \end{pmatrix}, S = \begin{pmatrix} 1 & 0 \\ 0 & i \end{pmatrix}, T = \begin{pmatrix} 1 & 0 \\ 0 & e^{\frac{i\pi}{4}} \end{pmatrix}, R_Z(\theta) = \begin{pmatrix} e^{-\frac{i\theta}{2}} & 0 \\ 0 & e^{\frac{i\theta}{2}} \end{pmatrix}, \qquad Ph(\theta) = \begin{pmatrix} e^{i\theta} & 0 \\ 0 & e^{i\theta} \end{pmatrix}.$$

We use $U_i$ to denote the application of gate U to the $i$-th qubit. In linear algebra, the application of a gate to a state $(a_0, a_1, \ldots, a_{2^n-1})^T$ corresponds to the matrix multiplication $(I_{i-1} \otimes U \otimes I_{n-i}) \cdot (a_0, a_1, \ldots, a_{2^n-1})^T$, where $I_j$ is the $2^j$ dimensional identity matrix and $\otimes$ is the *tensor product*. Under the tree view, a gate operation corresponds to a tree transformation on every two neighboring subtrees at level $i$. We use Fig. 7 to illustrate how the transformation $U_3$ works. Here $T_0$ and $T_1$ are the 0-subtree and 1-subtree of a node labeled $x_3$. Their leaves encode the amplitudes $a_1, a_2, \ldots, a_k$ and $b_1, b_2, \ldots, b_k$, respectively. After applying $U_3$, the leaves of $T_0$ become $(u_1 \cdot a_1 + u_2 \cdot b_1), (u_1 \cdot a_2 + u_2 \cdot b_2), \ldots, (u_1 \cdot a_k + u_2 \cdot b_k)$, obtained by multiplying the amplitudes of $T_0$ with $u_1$, those of $T_1$ with $u_2$, and summing up the two. Similarly, the leaves of $T_1$ become $(u_3 \cdot a_1 + u_4 \cdot b_1), (u_3 \cdot a_2 + u_4 \cdot b_2), \ldots, (u_3 \cdot a_k + u_4 \cdot b_k)$. The same transformation occurs in all neighboring subtrees at the same level. It is essential to note that the sum of probabilities remains the same after the U gate application, as U is unitary. We provide examples of applying $X_2$, $Z_2$, $H_1$ on $q$ in Fig. 8b, Fig. 8c, and Fig. 8d, respectively.

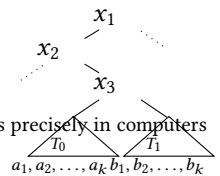

Fig. 7.  $T_0$ and $T_1$

---

[3]Note that we do not discuss how to represent complex numbers; representing complex numbers precisely in computers is an orthogonal issue to our work and can be handled by, e.g., the approach from [63] and [50].

### 2.2.2 Two frequently used sub-categories.

In fact, most of the quantum gates have a simpler structure than the general case. Except the $R_X(\theta)$ and H gates, all other considered single-qubit gates belong to the following two categories (or their composition): (1) the X (negation) gate and (2) diagonal matrix ($u_2 = u_3 = 0$) gates. This allows more efficient automata algorithms than the general case.

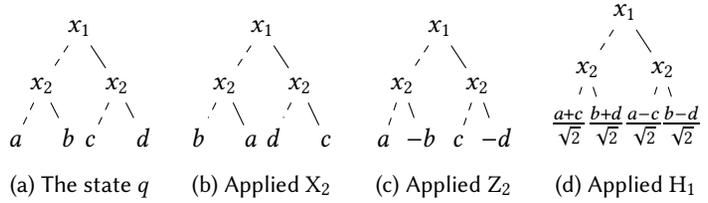

(a) The state $q$    (b) Applied $X_2$    (c) Applied $Z_2$    (d) Applied $H_1$

Fig. 8. The effect of applying single-qubit quantum gates to the state $q$.

The X gate is the quantum "negation" gate. Applying $X_i$ on a quantum state effectively swaps the 0- and 1-branches of all nodes at the level $i$. An example of applying $X_2$ on $q$ is available at Fig. 8b.

Gates with diagonal matrices, e.g., Z, S, T, $R_Z(\theta)$, and $Ph(\theta)$, multiply all 0- or 1-subtrees under nodes labeled $x_i$ by the complex values $r_0$ and $r_1$, respectively. We note that $|r_0|$ and $|r_1|$ always equal 1. We use $D^{r_0, r_1}$ to denote a gate with the diagonal matrix $D^{r_0, r_1} = \begin{pmatrix} r_0 & 0 \\ 0 & r_1 \end{pmatrix}$. We refer the reader to Fig. 8c for an example of the application of $Z_2 = D_2^{1, -1}$ on $q$. Gates with anti-diagonal matrices, e.g., Y, can be composed from X and $D^{i, -i}$ (i.e., $Y = X \cdot D^{i, -i}$).

### 2.2.3 Controlled gates.

A controlled gate CU uses another quantum gate U as its parameter. CU has a control qubit $x_c$ and the gate U is applied only when the control qubit $x_c$ has value 1. The controlled X gate $CX_2^1$ has the control qubit $x_1$ and would apply $X_2$ when $x_1$ is valued 1. The result of applying $CX_2^1$ on $q$ is available in Fig. 9b. Observe that $X_2$ is only applied to the 1-subtree of $x_1$. On the other hand, the result of applying $CX_1^2$ on $q$ is available in Fig. 9c. Observe that all 0-subtrees at level 2 remain the same as in $q$, but the 1-subtrees at level 2 are updated to the corresponding ones after applying the $X_1$ gate to $q$.

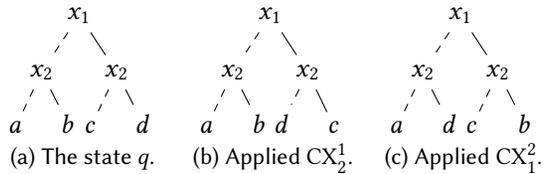

(a) The state $q$.    (b) Applied $CX_2^1$.    (c) Applied $CX_1^2$.

Fig. 9. The effect of applying controlled gates to the state $q$.

### 2.2.4 Quantum circuits.

As we mentioned before, a quantum circuit is a sequence of quantum gates. Executing a circuit effectively performs a sequence of tree updates following the gates' semantics. We often represent a quantum circuit using a diagram as in Fig. 1, which is also written as $H_1 CX_2^1$.

As an analogy, in classical circuits, a state corresponds to a computational basis state, and a gate application transforms one basis state to another. This process can be represented using a single tree branch. In contrast, encoding quantum states requires accounting for the amplitude values of all computational basis states, which necessitates using a complete tree structure for representation. Our method captures key quantum-specific features, particularly the ability to encode superpositions of basis states and their associated amplitudes. While our methods can be adapted to verify classical circuits by simplifying the state representation and adding support to classical gates, the primary distinction lies in our approach to handling quantum superposition and gate semantics—features that are absent in classical computation.

## 3   LEVEL-SYNCHRONIZED TREE AUTOMATA

In this paper, a new tree automata model called *Level-Synchronized Tree Automata* (LSTAs) is developed. The expressiveness of this model is incomparable with the traditional tree automata model (Theorem 17) while maintaining important properties such as being closed under union and intersection. It also allows for testing language emptiness and inclusion, enabling the testing of whether all reachable states are included in the post-condition. One crucial advantage of LSTAs is that they annotate transitions with "choices" and use them to coordinate between tree branches, enabling efficient quantum state encoding and gate operations (see Section 2.2).

### 3.1   Formal Definition of LSTAs

*Binary Trees.* We use $\mathbb{N}$ to represent the set of natural numbers (without 0), $\mathbb{N}_0 = \mathbb{N} \cup \{0\}$ to represent the set of non-negative integers, and $\mathbb{B} = \{0, 1\}$ to represent the Boolean values. A ranked alphabet is a set of symbols $\Sigma$ with a corresponding rank given by a function $\#\colon \Sigma \to \{0, 2\}$. The symbols with rank 0 are called *leaf* symbols, and those with rank 2 are called *internal* symbols.

A *binary tree* is a finite map $T\colon \{0, 1\}^* \to \Sigma$ that maps *tree nodes* (i.e., words over the alphabet $\{0, 1\}$) to symbols in $\Sigma$, and satisfies that (1) the domain of $T$ is *prefix-closed* and (2) if $T(v) = f$ and $f$ is internal, then the set of *children* of $v$ in $T$ is $\{v.0, v.1\}$, and if $f$ is a leaf symbol then $v$ has no children. Nodes labeled by leaf symbols and internal symbols are called *leaf nodes* and *internal nodes*, respectively. A node's *height* is its word length, denoted $\mathrm{ht}(w)$; e.g., $\mathrm{ht}(01010) = 5$. A node $w$ is at tree level $i$ when $\mathrm{ht}(w) = i$. A tree is *perfect* if all leaf nodes have the same height. We need only *perfect binary trees* to represent quantum states or vectors of sizes $2^n$.

*Example 4.* The quantum state of Fig. 2 corresponds to a tree $T$ with $T(\epsilon) = x_1, T(0) = T(1) = x_2,$ $T(00) = T(11) = 0,$ and $T(01) = T(10) = \frac{1}{\sqrt{2}}$, where $\epsilon$ is an empty string. We have $\mathrm{dom}(T) = \{\epsilon, 0, 1, 00, 01, 10, 11\}$. Children of the node 0 are 00 and 01. The leaf node 01 has no children.

*Definition 5.* A *level-synchronized tree automaton* (LSTA) is a tuple $\mathcal{A} = \langle Q, \Sigma, \Delta, \mathcal{R} \rangle$ where

(1) $Q$ is a finite set of *states*, $\mathcal{R} \subseteq Q$ is a set of *root states*, and $\Sigma$ is a ranked alphabet.
(2) $\Delta$ is a set of transitions of the form $q \ -C\!\!\to f(q_1, q_2)$ (*internal trans.*) or $q \ -C\!\!\to f$ (*leaf trans.*), where $C \subseteq \mathbb{N}_0$ is a finite set of *choices* (represented as natural numbers), $q, q_1, q_2 \in Q$, and $f \in \Sigma$. In figures, we draw internal and leaf transitions as 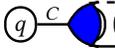 and 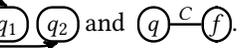.
(3) We call $q$, $f$, $C$, and $\{q_1, q_2\}$ the *top*, the *symbol*, the *choices*, and the *bottom*, respectively, of the transition $\delta$, and denote them by $\mathrm{top}(\delta)$, $\mathrm{sym}(\delta)$, $\ell(\delta)$, and $\mathrm{bot}(\delta)$, respectively. We use $|\mathcal{A}|$ to denote $\mathcal{A}$'s number of states.
(4) We further require that the choices of transitions with the same top state are disjoint, i.e., $\forall \delta_1 \neq \delta_2 \in \Delta\colon \mathrm{top}(\delta_1) = \mathrm{top}(\delta_2) \implies \ell(\delta_1) \cap \ell(\delta_2) = \emptyset$.

*The language of an LSTA.* A *run* of an LSTA $\mathcal{A}$ on a tree $T$ is a total map $\rho\colon \mathrm{dom}(T) \to \Delta$ from tree nodes to transitions of $\mathcal{A}$ such that for each node $v \in \mathrm{dom}(T)$, when $v$ is an internal node, $\rho(v)$ is of the form $q \ -C\!\!\to T(v)(q_0, q_1)$, where the two bottom states $q_0 = \mathrm{top}(\rho(v.0))$ and $q_1 = \mathrm{top}(\rho(v.1))$ are the two top states of $v$'s children $\rho(v.0)$ and $\rho(v.1)$. When $v$ is a leaf node, $\rho(v)$ is of the form $q \ -C\!\!\to T(v)$.

*Example 6.* We define a *run* $\rho$ of the LSTA $\mathcal{A}$ in Fig. 3 on the tree $T$ in Fig. 2 as follows. We have

$$\rho(\epsilon) = p \dashv 1\!\!\mapsto x_1(q_+, q_\pm), \qquad \rho(0) = q_+ \dashv 2\!\!\mapsto x_2(r_0, r_+), \qquad \rho(1) = q_\pm \dashv 2\!\!\mapsto x_2(r_\pm, r_0),$$

$$\rho(01) = r_+ \dashv 1,2\!\!\mapsto \tfrac{1}{\sqrt{2}}, \qquad \rho(10) = r_\pm \dashv 1\!\!\mapsto \tfrac{1}{\sqrt{2}}, \qquad \rho(00) = \rho(11) = r_0 \dashv 1,2\!\!\mapsto 0. \qquad \square$$

We define the *level d* of a run $\rho$ as the set of transitions with height $d$

$$\texttt{level}(\rho, d) := \{\rho(w) \mid w \in \mathrm{dom}(T) \wedge \mathrm{ht}(w) = d\}.$$

The run $\rho$ is *accepting* if $\mathrm{top}(\rho(\epsilon)) \in \mathcal{R}$ and all transitions from the same level share some common choice, i.e., $\forall d \in \mathbb{N}$: $\bigcap_{\delta \in \texttt{level}(\rho, d)} \ell(\delta) \neq \emptyset$—in other words, transitions at each tree level are *synchronized*. The *language* of $\mathcal{A}$ is the set $\mathcal{L}(\mathcal{A})$ of trees $T$ with an accepting run.

*Example 7.* Continuing from Example 6, we have $\texttt{level}(\rho, 1) = \{q_+ \dashv 2\!\!\mapsto x_2(r_0, r_+), q_\pm \dashv 2\!\!\mapsto x_2(r_\pm, r_0)\}$ and $\texttt{level}(\rho, 2) = \{r_+ \dashv 1,2\!\!\mapsto \tfrac{1}{\sqrt{2}}, r_0 \dashv 1,2\!\!\mapsto 0, r_\pm \dashv 1\!\!\mapsto \tfrac{1}{\sqrt{2}}\}$. Observe that $\rho$ is accepting because $\mathrm{top}(\rho(\epsilon)) = p \in \mathcal{R}$, the transitions from $\texttt{level}(\rho, 1)$ have a common choice 2, and those from $\texttt{level}(\rho, 2)$ have a common choice 1. $\qquad \square$

We defer a detailed discussion on properties of LSTAs to Section 6. At this point, we just note that the *inclusion test* over LSTAs, i.e., checking if $\mathcal{L}(\mathcal{A}_1) \subseteq \mathcal{L}(\mathcal{A}_2)$ for LSTAs $\mathcal{A}_1$ and $\mathcal{A}_2$, is decidable.

## 4  USING LSTAS TO DESCRIBE CORRECTNESS PROPERTIES

We can use an LSTA to encode a set of perfect binary trees (quantum states or vectors) and use them as pre- and post-conditions for verification of quantum circuits. Below, we will provide examples of the verification problems and the corresponding specifications given using LSTAs.

### 4.1  Verification of Oracle-Based Algorithms

An *oracle circuit* is a black box circuit used to encode a specific function. It plays a crucial role in many quantum algorithms by providing a way to query information in a single computational step. In the case of Grover's search algorithm [29], the oracle circuit encodes a function $f(x)\colon \mathbb{B}^n \rightarrow \mathbb{B}$ that outputs 1 if $x$ is the solution and 0 otherwise. In the case of the Bernstein-Vazirani algorithm [8], the oracle circuit encodes a secret bit string. To verify the correctness of these algorithms against all possible oracles, one way is to create a *parameterized* oracle circuit that uses input qubits and control gates to generate the corresponding oracle circuit. By composing the parameterized oracle circuit and the circuit to be verified, we can create a framework to verify the correctness of these circuits against all oracles.

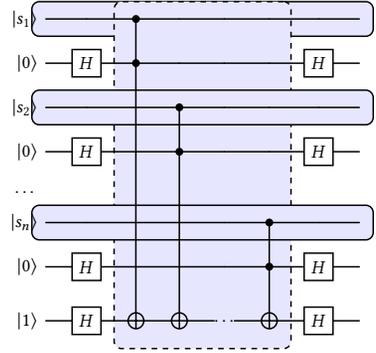

Fig. 10. BV circuit. Standard Toffoli gates have two $\bullet$ as controls and one $\oplus$ as the target.

Taking verification of the Bernstein-Vazirani algorithm (BV) as an example, the composed circuit consists of $2n+1$ qubits (Fig. 10), where the highlighted part 🟦 is the oracle circuit and the rest is the circuit under verification. We emphasize that in our setting, we consider parameterized oracle with the secret provided as a part of the input of the circuit. The qubits $s_1, s_2, \ldots, s_n$ serve as the input for the parameterized oracle circuit, and the other qubits act as the working tape of the circuit being verified, particularly, the last qubit acts as the ancilla, i.e., an *auxiliary variable* in the sense of classical program verification. We will verify the correctness of the implementation using the precondition $\{|s_1 0 s_2 0 \ldots s_n 0 1\rangle \mid s_1, s_2, \ldots s_n \in \mathbb{B}\}$ and the postcondition $\{|s_1 s_1 s_2 s_2 \ldots s_n s_n 1\rangle \mid s_1, s_2, \ldots s_n \in \mathbb{B}\}$. That is, considering all possible secret strings $s_1 s_2 \ldots s_n$ as the input of the oracle circuit and verifying that the BV circuit finds the same string at the output. We show the LSTA representing the postcondition in Fig. 11. In this LSTA, the states $q_0^i$ generate subtrees with leaves 0. From the state $q^i$, the LSTA picks a secret bit $b$ using the disjunctive branch, remembers it in states $q_R^{i+1}$ ($b = 1$) or $q_L^{i+1}$ ($b = 0$), and repeats the value in the next transition. The precondition can be modeled in a similar manner to Fig. 11 and, hence, omitted.

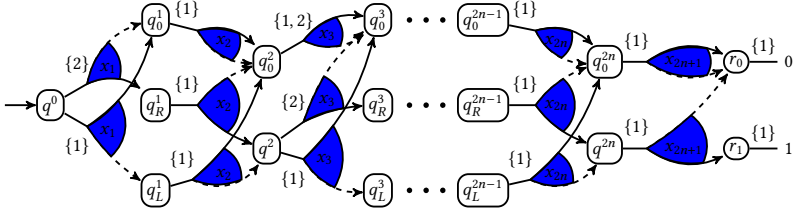

Fig. 11. An LSTA encoding the BV postcondition $|s_1s_1 \ldots s_n s_n 1\rangle$

## 4.2    Verification of Compound Multi-Control Quantum Gates

When implementing quantum algorithms on quantum computers, which have a limited set of supported gates, one often needs to find a way how to implement an unsupported gate by composing several natively supported gates. For instance, the multi-control Toffoli gate is not typically supported and is often created using standard Toffoli gates (cf. Fig. 12). In this type of circuit, qubits $|c_1\rangle, \ldots, |c_n\rangle$ serve as the control, those $|0\rangle$ below the controlled qubits are the ancilla qubits, and the last qubit $|t\rangle$ is the target. For this circuit to be correct, it should hold that (1) it maps ancilla qubits $|0^{n-1}\rangle$ to $|0^{n-1}\rangle$ (we do not impose any restrictions on the behavior of the circuit if the input ancillas are not $|0^n\rangle$) and (2) the operation of the circuit on the rest of the qubits is equivalent to a multi-control Toffoli gate, i.e., for computational bases of the form $|c_1 \ldots c_n t\rangle$, it swaps the amplitudes of bases $|1^n 0\rangle$ and $|1^n 1\rangle$ and keeps all the other bases the same.

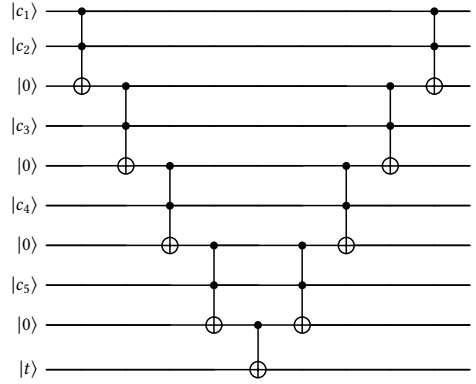

Fig. 12. Multi-control Toffoli with 5 control qubits.

A fundamental issue with specifying the functionality of such circuits using Hoare triples is that we need to express a mapping between input and output quantum states, which is not directly expressible using only sets of states[4]. In the case of multi-control gates, we can use the fact that the values of the control qubits should not change at the output, and we only care about the case the values of the ancillas remain 0, the only qubit whose value will change is the target $|t\rangle$. Hence, we reduce the verification problem to two sub-problems against the two pairs of pre- and postconditions $\text{Pre}_k$ and $\text{Post}_k$ below, for the value of $|t\rangle$ being $k \in \mathbb{B}$.

$$\text{Pre}_k = \{|c_1 c_2 0 c_3 0 \ldots c_n 0 k\rangle \mid c_1, \ldots, c_n \in \mathbb{B}\}$$
$$\text{Post}_k = \{|c_1 c_2 0 c_3 0 \ldots c_n 0 k'\rangle \mid c_1, \ldots, c_n \in \mathbb{B}, k' = (c_1 \wedge \ldots \wedge c_n) \oplus k\}$$

---

[4]One could, indeed, make a copy of the input qubit values, but this would cause a blow-up in the size of the underlying representation, losing the compactness of LSTAs.

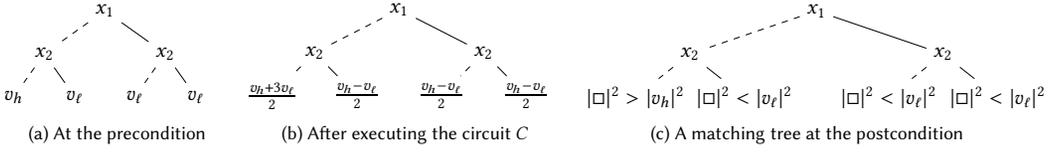

Fig. 14. Verification of a circuit $C$ amplifying the amplitude of $|00\rangle$.

Intuitively, the postcondition says that if all control qubits are set to 1, then the value of the target should be flipped (denoted using the xor operator $\oplus$). $\text{Pre}_k$ and $\text{Post}_k$ can be modelled using LSTAs in a similar way as in Fig. 11. Specification of other multi-control gates could be done likewise.

## 4.3 Equivalence Checking

When considering the LSTA model's expressiveness as a specification language, it is interesting to note that it can be used to express that a circuit implements the *identity* function, allowing for checking of circuit equivalence. More precisely, given two circuits $C_1$ and $C_2$, we can check if they are equivalent by checking if $C_1 C_2^\dagger$ is an identity, where $C_2^\dagger$ can be obtained from $C_2$ by reverting it (i.e., inputs are swapped with outputs) and substituting every gate by its inverse. Since all quantum gates correspond to unitary matrices, their inverses can be obtained by taking their conjugate transpose. For instance, the inverse of a single qubit gate $U = \left(\begin{smallmatrix} u_1 & u_2 \\ u_3 & u_4 \end{smallmatrix}\right)$ is the gate $U^\dagger = \left(\begin{smallmatrix} \overline{u_1} & \overline{u_3} \\ \overline{u_2} & \overline{u_4} \end{smallmatrix}\right)$ in the same form, where $\overline{u}$ is the complex conjugate of $u$ (therefore, whenever we can implement the gate $U$, we can also obtain an implementation of the gate $U^\dagger$).

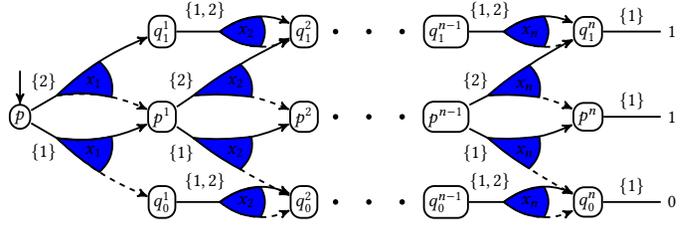

Fig. 13. LSTA for the pre- and post-condition for equivalence checking

We can then verify if an $n$-qubit circuit is an identity by testing it against $2^n$ linearly independent *vectors* and checking if each resulting vector matches its input vector (this is due to the fact that the identity matrix is the only matrix that maps all vectors back to themselves; due to linearity, it suffices to only try a maximal set of linearly independent vectors). Using LSTAs, all $2^n$ of these tests can be performed simultaneously. We use trees corresponding to the set of vectors $\{(0, 0, \ldots, 1), \ldots, (0, 1, \ldots, 1), (1, 1, \ldots, 1)\}$[5] simultaneously as both the precondition and the postcondition. The corresponding LSTA can be found in Fig. 13, with the number of transitions linear to the number of qubits. We note three key observations: (i) the set of vectors is linearly independent, (ii) the vectors do not represent quantum states, which is fine due to the linearity of quantum gate operations, and (iii) each vector has a unique number of ones; therefore, every vector has a distinct *Euclidean norm*. Since quantum gates are unitary operators, this norm is preserved, ensuring that there will be exactly one vector with the specified norm in the output. Note that not all linearly independent set of vectors can be used, such as the standard basis of $\mathbb{R}^{2^n}$, since the vectors there have the same norm, so we would lose the one-to-one correspondence between the input and output vectors.

## 4.4 Verification of Amplitude Constraints

Similarly to the work of [17], LSTAs can be extended to support symbolic amplitudes. For example, when verifying an amplitude amplification algorithm (such as Grover's search [29]), we can use variables $v_h$ and $v_l$ as amplitudes (leaf labels) of the LSTA representing the precondition and describe the relation between the two variables using a global constraint, such as $\varphi\colon |v_h + 3v_\ell| > |2v_h|$, which was used in [17] for Grover's search. A tree accepted by such an LSTA is in Fig. 14a and the output obtained after symbolically executing a circuit $C$ that amplifies the amplitude of $|00\rangle$ over the tree is in Fig. 14b. At the post-condition LSTA, we label the leaves with predicates in the form of $|\square|^2 > |x|^2$ or $|\square|^2 > 90\%$ to denote that the matching leaf probability (denoted as $|\square|^2$) should be bigger than the value of the precondition $|x|^2$ or the constant 90 %, respectively, for $x$ being some of the used variables. In Fig. 14c, we demonstrate an example of a tree accepted by the post-condition LSTA. Notice that if we put all leaf amplitudes of Fig. 14b into the matching $\square$ of the tree in Fig. 14c, the resulting formula is implied by the global constraint $\varphi$. For example, for the branch $|00\rangle$, we have that $|v_h + 3v_\ell| > |2v_h| \implies |\frac{v_h + 3v_\ell}{2}|^2 > |v_h|^2$ is valid. In such a case, we say that the resulting tree is accepted in the post-condition. The symbolic extension would allow us to verify a circuit against properties such as that the full Grover's circuit has >90 % probability of finding a correct answer or that an *amplitude amplification* algorithm increases the probability of finding a correct answer in each iteration.

---

[5]The set can be defined formally as the set $\{(b_1, b_2, \ldots, b_{2^n}) \in \{0, 1\}^{2^n} \mid 0 \le b_1 \le b_2 \le \ldots \le b_{2^n} = 1\}$.

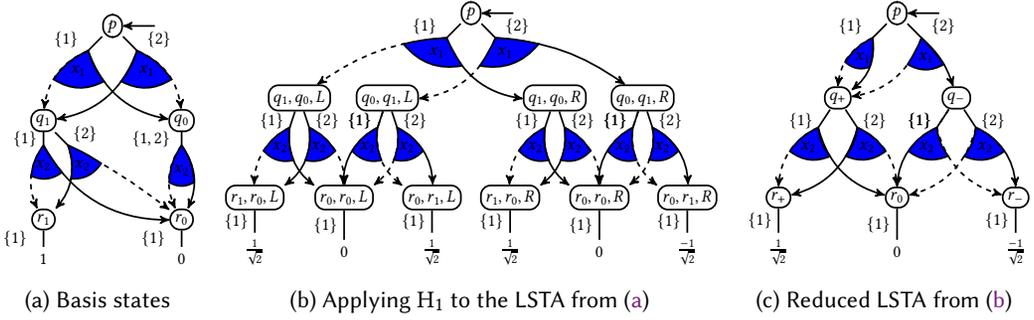

(a) Basis states          (b) Applying H$_1$ to the LSTA from (a)          (c) Reduced LSTA from (b)

Fig. 15.  An example of applying a single-qubit gate on a set of quantum states represented using an LSTA.

## 5   QUANTUM GATES OPERATIONS

Assuming that preconditions and postconditions are given as LSTAs, our next step in the verification of a quantum circuit is to compute the set of states reachable from the precondition after executing the circuit. In this section, we will show, given an LSTA $\mathcal{A}$ representing a set of quantum states and a quantum gate U, how to construct another LSTA $\mathrm{U}(\mathcal{A})$ with $\mathcal{L}(\mathrm{U}(\mathcal{A})) = \{\mathrm{U}(T) \mid T \in \mathcal{L}(\mathcal{A})\}$. Applying the construction for each gate in the circuit, we will then obtain an LSTA representing the set of reachable quantum states.

### 5.1   General Single-Qubit Gate

Let $\mathcal{A} = \langle Q, \Sigma, \Delta, \mathcal{R} \rangle$ be an LSTA representing a set of quantum states and U be a single-qubit gate $\mathrm{U} = \begin{pmatrix} u_1 & u_2 \\ u_3 & u_4 \end{pmatrix}$. Recall that applying U to the $t$-th qubit of a quantum state combines the 0-subtree $T_0$ and the 1-subtree $T_1$ under each node labelled with $x_t$. For every pair of $T_0$ and $T_1$ under a node labeled $x_t$, with leaf amplitudes of $a_1, a_2, \ldots, a_k$ and $b_1, b_2, \ldots, b_k$ respectively, the new state will have a new 0-subtree with the amplitudes $(u_1 \cdot a_1 + u_2 \cdot b_1), (u_1 \cdot a_2 + u_2 \cdot b_2), \ldots, (u_1 \cdot a_k + u_2 \cdot b_k)$, and a 1-subtree with the amplitudes $(u_3 \cdot a_1 + u_4 \cdot b_1), (u_3 \cdot a_2 + u_4 \cdot b_2), \ldots, (u_3 \cdot a_k + u_4 \cdot b_k)$. The construction is lifted from trees to LSTAs in Algorithm 1. The algorithm first constructs the transitions of $\mathrm{U}_t(\mathcal{A})$, which are partitioned into the following four sets:

- $\Delta'_{<t}$ indicates that the transitions of qubits before $x_t$ remain the same.

---

**Algorithm 1:** Application of a single-qubit gate on an LSTA

---

**Input:** An LSTA $\mathcal{A} = \langle Q, \Sigma, \Delta, \mathcal{R} \rangle$ and a gate $U_t = \left( \begin{smallmatrix} u_1 & u_2 \\ u_3 & u_4 \end{smallmatrix} \right)$

**Output:** $U_t(\mathcal{A})$

1   $\Delta'_{<t} := \{q \xrightarrow{-C} x_i(q_l, q_r) \in \Delta \mid i < t\};$

2   $\Delta'_{=t} := \{q \xrightarrow{-C} x_t((q_l, q_r, L), (q_l, q_r, R)) \mid q \xrightarrow{-C} x_t(q_l, q_r) \in \Delta\};$

3   $\Delta'_{>t} := \{(q_l, q_r, D) \xrightarrow{-C_1 \cap C_2} x_i((q_l^l, q_r^l, D), (q_l^r, q_r^r, D)) \mid i > t \wedge q_l \xrightarrow{-C_1} x_i(q_l^l, q_l^r), q_r \xrightarrow{-C_2} x_i(q_r^l, q_r^r) \in \Delta, D \in \{L, R\}\};$

4   $\Delta'_0 := \{(q_l, q_r, L) \xrightarrow{-C_1 \cap C_2} u_1 \cdot a + u_2 \cdot b, (q_l, q_r, R) \xrightarrow{-C_1 \cap C_2} u_3 \cdot a + u_4 \cdot b \mid q_l \xrightarrow{-C_1} a, q_r \xrightarrow{-C_2} b \in \Delta\};$

5   $Q' := Q \cup (Q \times Q \times \{L, R\}); \Delta' := \Delta'_{<t} \cup \Delta'_{=t} \cup \Delta'_{>t} \cup \Delta'_0; \Sigma' := \Sigma \cup \{u_1 \cdot a + u_2 \cdot b, u_3 \cdot a + u_4 \cdot b \mid a, b \in \Sigma_0\};$

6   **return** $\langle Q', \Sigma', \Delta', \mathcal{R} \rangle;$

---

- $\Delta'_{=t}$ initiates a product construction, where both the left-hand side state $q_l$ and the right-hand side state $q_r$ operate simultaneously. The symbols $L$ and $R$ are used to remember the operation to perform when the construction reaches the leaves; $L$-labeled states combine leaf values using $u_1$ and $u_2$ while $R$-labeled ones use $u_3$ and $u_4$.
- $\Delta'_{>t}$ continues the product construction while remembering $L$ and $R$ and taking the intersection of the choices from both sides.
- $\Delta'_0$ combines the probability amplitude of the leaves based on the symbol $L$ and $R$.

The H gate is a special case of a single-qubit gate, with $u_1 = u_2 = u_3 = \frac{1}{\sqrt{2}}$, and $u_4 = -\frac{1}{\sqrt{2}}$. An example of applying the H gate to an LSTA can be found in Fig. 15, and the result of applying our LSTA reduction algorithm (Section 6.3) to simplify its structure further can be found in Fig. 15c.

**Theorem 8.** $\mathcal{L}(U_t(\mathcal{A})) = \{U_t(T) \mid T \in \mathcal{L}(\mathcal{A})\}$ *and* $|U_t(\mathcal{A})| = O(|\mathcal{A}|^2).$

## 5.2 Controlled Gate

For simplification, we will focus on the controlled gate $CU_t^c$. This gate applies a single-qubit gate $U_t$ when the control qubit $x_c$ is 1. We will be using the LSTA $\mathcal{A} = \langle Q, \Sigma, \Delta, \mathcal{R} \rangle$ as the input.

While constructing $CU_t^c(\mathcal{A})$, when we encounter a transition labeled with the control qubit $x_c$, on the 0-subtree, we want to simulate the behavior of $\mathcal{A}$ (because nothing should change when the control qubit is 0), while on the 1-subtree, we want to simulate the LSTA $U_t(\mathcal{A})$. We, however, also need to keep the synchronization between the 0- and 1-subtrees in order not to mix the 0-subtree of one quantum state with the 1-subtree of another quantum state from $\mathcal{L}(\mathcal{A})$.

Our algorithm is built on top of Algorithm 1, which computes the LSTA $U_t(\mathcal{A})$. The main benefit of having $U_t(\mathcal{A})$ is that the $x_c$-labeled transitions in $U_t(\mathcal{A})$ have the information from both $\mathcal{A}$ and $U_t(\mathcal{A})$ on both 0- and 1- subtrees stored in product states (states from $Q \times Q \times \{L, R\}$). Therefore, we only need to adjust its 0-subtree to stay in $\mathcal{A}$.

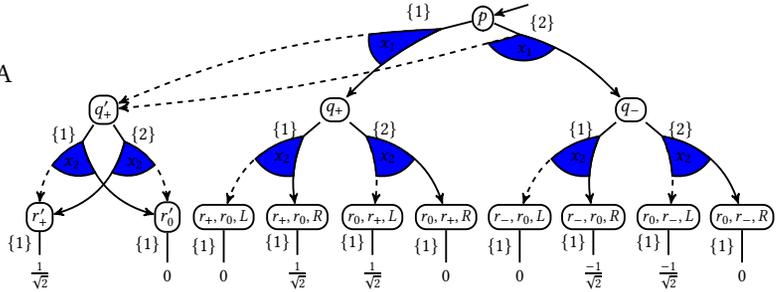

Fig. 16. An LSTA obtained after applying $CX_2^1$ to the LSTA in Fig. 15c

We present the full construction in Algorithm 2. The algorithm updates transitions from $U_t(\mathcal{A})$ labeled with the control qubit $x_c$ such that the 0-branch will connect to the corresponding state in the transition system of $\mathcal{A}'$, a primed copy of the input LSTA (Lines 3–6). For the case of $c < t$, we simply redirect the 0-branch from the state $q_1$ to its primed version $q'_1$. When $c > t$, $q_1$ is a product state of the form $(q_a, q_b, D)$ for $D \in \{L, R\}$, so we reconnect $q_1$ to $q'_a$ for the $L$ case and to $q'_b$ for

---

**Algorithm 2:** Application of a controlled gate on an LSTA

---

**Input:** An LSTA $\mathcal{A} = \langle Q, \Sigma, \Delta, \mathcal{R} \rangle$, a single-qubit gate $U_t = \left( \begin{smallmatrix} u_1 & u_2 \\ u_3 & u_4 \end{smallmatrix} \right)$, a control qubit $x_c$
**Output:** $CU_t^c(\mathcal{A})$

1 Build $U_t(\mathcal{A}) = \langle Q^U, \Sigma^U, \Delta^U, \mathcal{R} \rangle$ using Algorithm 1, with $Q^U = Q \cup (Q \times Q \times \{L, R\})$;
2 Build $\mathcal{A}' = \langle Q', \Sigma, \Delta', \mathcal{R}' \rangle$, a primed copy of $\mathcal{A}$;
3 **foreach** $\delta = q \; -C \rightarrow x_c(q_1, q_2) \in \Delta^U$ **do**
4      **if** $q_1 \in Q$ **then** replace $\delta$ with $q \; -C \rightarrow x_c(q_1', q_2)$ in $\Delta^U$;             // case $c < t$
5      **else if** $q_1 = (q_a, q_b, L)$ **then** replace $\delta$ with $q \; -C \rightarrow x_c(q_a', q_2)$ in $\Delta^U$;
6      **else if** $q_1 = (q_a, q_b, R)$ **then** replace $\delta$ with $q \; -C \rightarrow x_c(q_b', q_2)$ in $\Delta^U$;
7 **return** $\langle Q^U \cup Q', \Sigma^U \cup \Sigma, \Delta^U \cup \Delta', \mathcal{R} \rangle$

---

the $R$ case. The construction keeps all other transitions intact. We provide an example in Fig. 16. We can generalize it to multi-control gates by allowing more controlled qubits at Line 3.

THEOREM 9. $\mathcal{L}(CU_t^c(\mathcal{A})) = \{CU_t^c(T) \mid T \in \mathcal{L}(\mathcal{A})\}$ and $|CU_t^c(\mathcal{A})| = |\mathcal{A}| + |U_t(\mathcal{A})|$.

### 5.3 Optimizations for the X Gates and Diagonal Matrix Gates

Most single-qubit gates implemented in state-of-the-art quantum computers belong to this category, making it worth considering for special treatment. Note that these constructions are similar to the *permutation-based* construction in [18], but differ in the need for the treatment of choices.

We first show the construction of the LSTA for the $X_t$ gate. For an LSTA $\mathcal{A}$, we can capture the effect of the $X_t$ gate to all quantum states in $\mathcal{L}(\mathcal{A})$ by swapping the left and the right children of all $x_t$-labeled transitions $q \; -C \rightarrow x_t(q_0, q_1)$, i.e., update them to $q \; -C \rightarrow x_t(q_1, q_0)$. All other transitions will stay the same. We use $X_t(\mathcal{A})$ to denote the LSTA constructed following this procedure.

THEOREM 10. $\mathcal{L}(X_t(\mathcal{A})) = \{X_t(T) \mid T \in \mathcal{L}(\mathcal{A})\}$ and $|X_t(\mathcal{A})| = |\mathcal{A}|$.

Applying the diagonal matrix gate $D_t^{r_0, r_1}$ on the qubit $x_t$ multiplies all leaves of all 0-subtrees rooted in $x_t$ nodes with $r_0$ and leaves of 1-subtrees rooted therein with $r_1$. The construction is formally given in Algorithm 3. In the algorithm, internal $x_t$-transitions $q \; -C \rightarrow x_t(q_0, q_1)$ are modified to $q \; -C \rightarrow x_t(q_0, q_1')$, where $q_1'$ is a copy of $q_1$ in the primed version $\mathcal{A}'$ of $\mathcal{A}$. Then all leaves in $\mathcal{A}$ are multiplied by $r_0$ and all leaves in $\mathcal{A}'$ are multiplied by $r_1$.

THEOREM 11. $\mathcal{L}(D_t^{r_0, r_1}(\mathcal{A})) = \{D_t^{r_0, r_1}(T) \mid T \in \mathcal{L}(\mathcal{A})\}$ and $|D_t^{r_0, r_1}(\mathcal{A})| = 2|\mathcal{A}|$.

## 6 LSTA ALGORITHMS

With the algorithms of LSTA quantum gates operations, we can now compute the set of reachable states from the precondition, represented as LSTA. We can then check if all reachable states are allowed by the postcondition, using the language inclusion algorithm we are going to present in

---

**Algorithm 3:** Application of a diagonal matrix gate on an LSTA

---

**Input:** An LSTA $\mathcal{A} = \langle Q, \Sigma, \Delta, \mathcal{R} \rangle$, a diagonal matrix gate $D_t^{r_0, r_1}$
**Output:** $D_t^{r_0, r_1}(\mathcal{A})$

1 Build $\mathcal{A}' = \langle Q', \Sigma, \Delta', \mathcal{R}' \rangle$, a primed copy of $\mathcal{A}$;
2 Replace internal $x_t$-transitions $q \; -C \rightarrow x_t(q_0, q_1)$ from $\Delta$ with $q \; -C \rightarrow x_t(q_0, q_1')$;
3 Replace leaf transitions $q \; -C \rightarrow k$ from $\Delta$ with $q \; -C \rightarrow r_0 \cdot k$; Add $r_0 \cdot k$ to $\Sigma$;
4 Replace leaf transitions $q' \; -C \rightarrow k$ from $\Delta'$ with $q' \; -C \rightarrow r_1 \cdot k$; Add $r_1 \cdot k$ to $\Sigma$;
5 **return** $\langle Q \cup Q', \Sigma, \Delta \cup \Delta', \mathcal{R} \rangle$;

---

---

**Algorithm 4:** LSTA non-emptiness algorithm

---

**Input:** An LSTA $\mathcal{A} = \langle Q, \Sigma, \Delta, \mathcal{R} \rangle$ with $\mathcal{R} \neq \emptyset$
**Output:** true iff $\mathcal{L}(\mathcal{A}) \neq \emptyset$

1  Guess $r \in \mathcal{R}$ and assign $S \leftarrow \{r\}$;
2  **while** $S \neq \emptyset$ **do**
3      Let $\Gamma \subseteq \Delta$ be a nondeterministically constructed set of transitions such that for all $q \in S$, there is exactly one transition $\delta_q \in \Gamma$ such that $\text{top}(\delta_q) = q$;
4      **if** $\Gamma$ *does not exist or* $\bigcap_{\delta \in \Gamma} \ell(\delta) = \emptyset$ **then return false**;
5      $S := \{q_1, q_2 \mid q \: \text{-}C\text{→} \: f(q_1, q_2) \in \Gamma\}$;
6  **return true**;

---

this section. Besides language inclusion, we will cover other decision problems, complexity, and algorithms of LSTA in this section for a complete presentation.

## 6.1 Intersection, Union, Complementation, and Emptiness Testing

THEOREM 12. *LSTAs are closed under union and intersection but not closed under complementation.*

PROOF. ⟨Complementation⟩ Fix a set of ranked alphabet $\Sigma = \{x, a\}$, we can construct an LSTA accepting empty language, however, we will show in Theorem 17 that we cannot construct one accepting all possible trees using $\Sigma$.

⟨Union⟩ Let $\mathcal{A}^j = \langle Q^j, \Sigma, \Delta^j, \mathcal{R}^j \rangle$, for $j = 1, 2$, be two LSTAs, their union $\mathcal{A}_1 \cup \mathcal{A}_2$ can be constructed by combining the transition systems of $\mathcal{A}_1$ and $\mathcal{A}_2$. This involves creating a disjoint union of their states, transitions, root states, and symbols. That is,

$$\mathcal{A}_{1 \cup 2} = \langle Q, \Sigma, \Delta, \mathcal{R} \rangle, \text{ where } Q = Q^1 \uplus Q^2, \mathcal{R} = \mathcal{R}^1 \uplus \mathcal{R}^2, \text{ and } \Delta = \Delta^1 \uplus \Delta^2.$$

⟨Intersection⟩ On the other hand, their intersection can be constructed via a *product construction* and reassign each pair of choices to a fresh number. That is,

$$\mathcal{A}_{1 \cap 2} = \langle Q', \Sigma, \Delta', \mathcal{R}' \rangle, \text{ where } Q' = Q^1 \times Q^2, \mathcal{R}' = \mathcal{R}^1 \times \mathcal{R}^2, \text{ and }$$

$$\Delta' = \{(q^1, q^2) \: \text{-}N(C^1 \times C^2)\text{→} \: f((q_l^1, q_l^2), (q_r^1, q_r^2)) \mid q^1 \: \text{-}C^1\text{→} \: f(q_l^1, q_r^1) \in \Delta^1 \wedge q^2 \: \text{-}C^2\text{→} \: f(q_l^2, q_r^2) \in \Delta^2\}$$

$$\cup \{(q^1, q^2) \: \text{-}N(C^1 \times C^2)\text{→} \: k \mid q^1 \: \text{-}C^1\text{→} \: k \in \Delta^1 \wedge q^2 \: \text{-}C^2\text{→} \: k \in \Delta^2\}$$

Here, $N(C^1 \times C^2)$ is a set of choices obtained by mapping every choice pair in $C^1 \times C^2$ to a unique number in $\mathbb{N}$. Proofs of the correctness of these constructions are standard. □

THEOREM 13. *LSTA emptiness is **PSPACE**-complete.*

PROOF. ⟨**PSPACE**-membership⟩ We will show that LSTA non-emptiness is in **NPSPACE**, which implies that LSTA emptiness is in **co-NPSPACE**, and the result will follow by Immerman–Szelepcsényi theorem (**co-NPSPACE = NPSPACE**) and Savitch's theorem (**NPSPACE = PSPACE**).

By Algorithm 4, we show that LSTA non-emptiness is in **NPSPACE**. At every step, Algorithm 4 only remembers the current value of $S$, $\Gamma$, and the new values of $S$, which are all of a polynomial size w.r.t. the size of $\mathcal{A}$. The algorithm takes advantage of the fact that at every level of a *LSTA* run, the transitions should take the same choice, and therefore all occurrences of $q$ at a level must agree on the same transition. As a result, the subtrees of the same state $q$ at a level are also the same, and, therefore, we need to keep track of only one occurrence.

**PSPACE-hardness** By reduction from the **PSPACE**-complete problem of universality of a *non-deterministic finite automaton* (NFA) that has in each state over every symbol exactly two non-deterministic transitions. One can show that universality for this sub-class of NFAs is still **PSPACE**-hard, e.g., by modifying the standard proof in [24, Theorem 3.13]. The proof in [24] goes by reduction from the **PSPACE**-complete problem of membership of a string in the language of a linearly bounded automaton (LBA). It constructs an NFA that rejects a sequence of configurations that corresponds to an accepting run of the LBA, and accepts all other sequences of symbols. So it holds that the input is not in the language of the LBA iff the language of the NFA is universal. This can be modified for the considered class of NFAs by relaxing the structure of the LBA's configurations, allowing to use "*empty symbols*" on the tape. Details are technical.

Let $\mathcal{M} = (Q, \Sigma, \delta, I, F)$ be an NFA of the class above over (standard unranked) alphabet $\Sigma$ with the sets of initial and final states $I, F \subseteq Q$ and the transition function $\delta\colon Q \times \Sigma \to 2^Q$, where it holds for all $q \in Q$ and $a \in \Sigma$ that $|\delta(q, a)| = 2$ as mentioned above. W.l.o.g., we assume that $\mathcal{M}$ contains at least one initial state and that $\Sigma = \{1, \ldots, n\}$. We construct the LSTA $\mathcal{A}_{\mathcal{M}} = (Q, \{\spadesuit, \circ\}, \Delta_\delta, I)$ where $\spadesuit$ and $\circ$ are new symbols with arity 0 and 2 respectively. The LSTA, intuitively, works as follows: Symbols from $\Sigma \cup \{0\}$ are encoded into choices on transitions. The tree generated by $\mathcal{A}_{\mathcal{M}}$ corresponds to the computational tree of the textbook algorithm performing universality check on $\mathcal{M}$ by doing on-the-fly determinization and checking for a macrostate that contains no state from $F$. $\mathcal{A}_{\mathcal{M}}$ will accept such a tree because only states that are non-accepting in $\mathcal{M}$ contain a leaf transition in $\mathcal{A}_{\mathcal{M}}$ (synchronized using choice 0). Formally, $\Delta_\delta$ is defined as follows:

$$\Delta_\delta = \{q \overset{0}{\dashrightarrow} \spadesuit \mid q \notin F\} \cup \{q \overset{i}{\dashrightarrow} \circ(q_1, q_2) \mid \delta(q, i) = \{q_1, q_2\}\}, \tag{1}$$

It holds that the (word) language of $\mathcal{M}$ is not $\Sigma^*$ iff the (tree) language of $\mathcal{A}_{\mathcal{M}}$ is non-empty.  □

## 6.2 Entailment Testing

**Theorem 14.** *The inclusion of languages of two LSTAs is decidable.*

**Proof.** Reduce to graph reachability The inclusion $\mathcal{L}(\mathcal{A}) \subseteq \mathcal{L}(\mathcal{B})$ between two LSTAs $\mathcal{A} = \langle Q_{\mathcal{A}}, \Sigma, \Delta_{\mathcal{A}}, \mathcal{R}_{\mathcal{A}} \rangle$ and $\mathcal{B} = \langle Q_{\mathcal{B}}, \Sigma, \Delta_{\mathcal{B}}, \mathcal{R}_{\mathcal{B}} \rangle$ can be reduced to graph reachability in a directed graph $(V, A)$, with the vertices in $V$ of the form $(D, \{F_1, \ldots, F_m\})$ where $D \subseteq Q_{\mathcal{A}}$ is the *domain* and $F_i\colon D \to 2^{Q_{\mathcal{B}}}$ is a total map that assigns sets of states of $\mathcal{B}$ to states of $\mathcal{A}$ from the domain $D$. The algorithm makes use of the following essential property of trees generated by an LSTA $\mathcal{A}$: if two nodes at the same level of a tree $T$ are labelled by the same state in an accepting run of $\mathcal{A}$ on $T$, then the subtrees rooted in these nodes are identical (this follows from the semantics of LSTAs and the restriction on transitions, cf. Section 3.1).

Intuitively, $D$ represents the set of states of $\mathcal{A}$ in a level of a run $\rho$ of $\mathcal{A}$, and every $F_i$ represents the same level of some possible run $\rho_i$ of $\mathcal{B}$ on the same tree, and how it can cover the run $\rho$ of $\mathcal{A}$. For instance, in Fig. 17, the state $q$ of $\mathcal{A}$ corresponds to the states $r$ and $s$ of $\mathcal{B}$ because they are used in the same tree level and the same tree nodes. So we have $F_i(q) = \{r, s\}$. Due to the property that all occurrences of a state at the same level in a run generate the same subtree mentioned above, we only need to maintain encountered states and their alignment with each another.

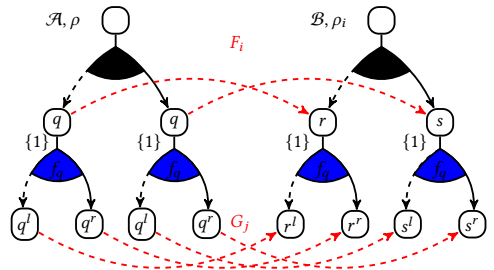

Fig. 17.  Computing $G_j$ from $F_i$

Source and terminal vertices They correspond to the root and leaf tree levels, respectively.

(1) A vertex $(\{q\}, \{\{q \mapsto r_1\}, \ldots, \{q \mapsto r_k\}\})$ with $q \in \mathcal{R}_{\mathcal{A}}$ and $\mathcal{R}_{\mathcal{B}} = \{r_1, \ldots r_k\}$ is a *source* vertex. Intuitively, both automata start their runs in their root states.

(2) A vertex $(\emptyset, \mathcal{F})$ where $\emptyset \notin \mathcal{F}$ is a *terminal* vertex. Intuitively, empty domain means that at that level of the run of $\mathcal{A}$, all branches of the tree have already ended at leaves, hence $\mathcal{A}$ accepts the tree. On the other hand, $\emptyset \notin \mathcal{F}$ means that $\mathcal{B}$ did not have any run on the same tree that would end at leaves and accept. The tree is therefore accepted only by $\mathcal{A}$.

⎡Arrows⎤ Starting from a vertex $\mathsf{U} = (D, \{F_1, \ldots, F_m\})$, we construct an arrow $(\mathsf{U}, \mathsf{V})$ with $\mathsf{V} = (E, \mathcal{G})$ as follows. First, we construct sets of transitions outgoing from $D$ such that in each set $\Gamma_{\mathcal{A}}$, we select exactly one downward transition $\delta_{q_{\mathcal{A}}}$ originating from each $q_{\mathcal{A}} \in D$, such that all transitions in $\Gamma_{\mathcal{A}}$ share a common choice (as required by the definition of an accepting run (cf. Section 3.1). Formally, given $D = \{q_1, \ldots, q_k\}$, we consider all sets of transitions $\Gamma_{\mathcal{A}} = \{\delta_1, \ldots, \delta_k\}$ such that the following formula holds:

$$(\forall 1 \leq i \leq k : \delta_i \in \Delta \wedge \mathsf{top}(\delta_i) = q_i) \quad \wedge \quad \bigcap \{\ell(\delta_i) \mid 1 \leq i \leq k\} \neq \emptyset. \tag{2}$$

The domain of the target $\mathsf{V}$ is then obtained from $\Gamma_{\mathcal{A}}$ as the set $E = \{q^l, q^r \mid q \, \text{-}C\text{→} \, f_q(q^l, q^r) \in \Gamma_{\mathcal{A}}\}$. If all transitions in $\Gamma_{\mathcal{A}}$ are leaf transitions, then $E = \emptyset$. We note that it is perfectly possible that no $\Gamma_{\mathcal{A}}$ exists for some vertex $\mathsf{U}$, in which case $\mathsf{U}$ has no successor.

For the construction of $\mathcal{G}$, we need to maintain the correspondences recorded in $F_1, \ldots, F_m$. Therefore, for every set $\Gamma_{\mathcal{A}}$ and every mapping $F_i \in \mathcal{F}$, we create a set of mappings $\{G_1, \ldots, G_n\}$ as follows: First, we construct sets $\Gamma_{\mathcal{B}}$ of (synchronized) transitions leaving $\mathsf{img}(F_i)$ in the same way as in Eq. (2) (with $D$ substituted by $\mathsf{img}(F_i)$). Then, for every transition $\delta_{\mathcal{A}} \in \Gamma_{\mathcal{A}}$, we find the corresponding transition $\delta_{\mathcal{B}}$ in the current $\Gamma_{\mathcal{B}}$ (i.e., such that $\mathsf{top}(\delta_{\mathcal{B}}) \in F_i(\mathsf{top}(\delta_{\mathcal{A}}))$). For the pair $\delta_{\mathcal{A}}$ and $\delta_{\mathcal{B}}$, we then check whether their symbols match, and if not, we dismiss the pair $\Gamma_{\mathcal{A}}$ and $\Gamma_{\mathcal{B}}$. On the other hand, if the symbols of all such pairs of transitions match, we construct the new state mapping $G_j$ such that for every $q \in E$, we set

$$G_j(q) = \big\{ s_q^l \mid r \, \text{-}C\text{→} \, f(q, q^r) \in \Gamma_{\mathcal{A}}, s \in F_i(r), s \, \text{-}C'\text{→} \, f(s_q^l, s_q^r) \in \Gamma_{\mathcal{B}} \big\} \cup$$
$$\big\{ s_q^r \mid r \, \text{-}C\text{→} \, f(q^l, q) \in \Gamma_{\mathcal{A}}, s \in F_i(r), s \, \text{-}C'\text{→} \, f(s_q^l, s_q^r) \in \Gamma_{\mathcal{B}} \big\}.$$

We then merge all sets $\{G_1, \ldots, G_n\}$ for all $F_i$'s into one set $\mathcal{G}$ and construct the vertex $(E, \mathcal{G})$.

The graph is finite because $Q_{\mathcal{A}}$ and $Q_{\mathcal{B}}$ are finite. The inclusion holds iff the graph has no path from a source to a terminal vertex.

⎡Witness of $\mathcal{L}(\mathcal{A}) \not\subseteq \mathcal{L}(\mathcal{B})$⎤ If we also remember the set $\Gamma_{\mathcal{A}}$ in the graph vertex, we can construct the tree to witness the non-inclusion following the path from a source vertex to a terminal vertex. Recall that the $D$ part of a vertex $(D, \{F_1, \ldots, F_m\})$ corresponds to a tree level of $\mathcal{A}$, and the set $\Gamma_{\mathcal{A}}$ gives the information on how the levels are linked together to form a tree.

**Theorem 15.** *The LSTA language inclusion problem is* **PSPACE**-*hard and in* **EXPSPACE**.

**Proof.** We can show that the problem is **PSPACE**-hard by a trivial reduction from the emptiness problem (Theorem 13). For the upper bound, let us analyze the complexity of the inclusion algorithm from the proof of Theorem 14. How many vertices are there in the graph? Since every node of the graph has the structure $(D, \{F_1, \ldots, F_m\})$, the set of vertices is included in the set $2^{Q_{\mathcal{A}}} \times 2^{Q_{\mathcal{A}} \to 2^{Q_{\mathcal{B}}}} = 2^{Q_{\mathcal{A}}} \times 2^{\left((2^{Q_{\mathcal{B}}})^{Q_{\mathcal{A}}}\right)}$ (we can change the original definition of $F_i$ to be a mapping of the type $F_i \colon Q_{\mathcal{A}} \to 2^{Q_{\mathcal{B}}}$). We can then bound the number of vertices by the number

$$2^{|Q_{\mathcal{A}}|} \cdot 2^{\left((2^{|Q_{\mathcal{B}}|})^{|Q_{\mathcal{A}}|}\right)} = 2^{|Q_{\mathcal{A}}|} \cdot 2^{\left(2^{|Q_{\mathcal{B}}| \cdot |Q_{\mathcal{A}}|}\right)} = 2^{|Q_{\mathcal{A}}| + \left(2^{|Q_{\mathcal{B}}| \cdot |Q_{\mathcal{A}}|}\right)} \tag{3}$$

The size of one node is then in $O(|Q_{\mathcal{A}}| + (2^{|Q_{\mathcal{B}}| \cdot |Q_{\mathcal{A}}|}))$, i.e., exponential. Since inclusion is checked by graph reachability, we can conclude that the problem is in **EXPSPACE**. □

---

**Algorithm 5:** LSTA reduction

---

**Input:** An LSTA $\mathcal{A} = \langle Q, \Sigma, \Delta, \mathcal{R} \rangle$
**Output:** A reduced LSTA with the same language as $\mathcal{A}$

1  *changed* := **true**;
2  **while** *changed* **do**
3      *changed* := **false**;
4      **foreach** $\delta_1 = q - C_1 \rightarrow x(B)$ *and* $\delta_2 = q - C_2 \rightarrow x(B) \in \Delta$ *s.t.* $C_1 \neq C_2$ *and* $B \in Q^0 \cup Q^2$ **do**
5         $\Delta := (\Delta \setminus \{\delta_1, \delta_2\}) \cup \{q - C_1 \cup C_2 \rightarrow x(B)\}$; *changed* := **true**;
6      **foreach** $p, q \in Q$ *s.t.* $p \neq q$ **do**
7         **if** $\forall B \in Q^0 \cup Q^2, C \subseteq \mathbb{N}_0, x \in \Sigma : q - C \rightarrow x(B) \in \Delta \Leftrightarrow p - C \rightarrow x(B) \in \Delta$ **then**
8            Merge $p$ and $q$; *changed* := **true**;

9  **return** $\mathcal{A}$;

---

## 6.3 LSTA Reduction

After executing corresponding operations for gates in the circuit on the precondition LSTA, the size of the LSTA may increase exponentially w.r.t. the number of gates, even though each gate operation is bounded by a quadratic factor. In practice, we employ a simple reduction algorithm for LSTAs (Algorithm 5), which significantly aids in controlling the size of the generated LSTAs. The algorithm effectively performs reduction w.r.t. the bottom-up bisimulation [1], i.e., it merges two states with the same downward-behavior (Lines 6–8). In addition, the algorithm also deals with transitions that are the same except for the set of choices—such transitions can be merged into one (Lines 4–5). An example of the output of a reduction can be found in Fig. 15c. When run on LSTAs that are acyclic (e.g., they represent quantum states with a fixed number of qubits), the algorithm can be optimized by considering states in an order starting from states that only have leaf symbols, going upwards, level by level.

## 6.4 Comparison with the Traditional Tree Automata Model

In traditional top-down deterministic tree automata (TAs), we would require the set of possible transitions leaving $q$ under $f$ to be of cardinality one. With LSTAs, we allow this set to be larger, but at every level of the run, the transitions should take the same choice. In contrast, a traditional TA is almost the same with LSTA, with the exception that it does not use the labeled choices for synchronization. Other modifications of traditional tree automata that allow global constraints to go beyond regularity has been considered in the literature [11, 26, 32, 33, 47]. However, none of them permits language inclusion test, which is essential for symbolic verification.

THEOREM 16 (SUCCINCTNESS). *Traditional TAs need at least $2^{n-1}$ transitions to encode all n-qubits GHZ states.*

PROOF. If we view each GHZ state as a tree (Fig. 5), there are $2^{n-1}$ 0-subtrees, and each 1-subtree corresponds to a unique 0-subtree. Using traditional tree automata without synchronization, we would require $2^{n-1}$ root transitions, each connecting to a unique pair of one 0-subtree and two 1-subtrees. If there were fewer root transitions, then there would be two different 0-subtrees generated by the same root transition, and any generated 1-subtree could be paired with both of them. This contradicts the condition that each 1-subtree corresponds to a unique 0-subtree. □

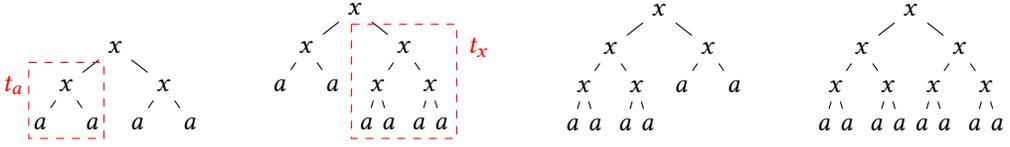

Fig. 19. Trees of $n = 1$ have $2^2$ varieties as branches below level 1

General quantum gates create a correspondence between the subtrees (cf. Section 2.2.1 and Fig. 7), and, hence, level-synchronization is critical in representing such sets of quantum states concisely.

**THEOREM 17.** *The class of languages recognized by LSTAs is incomparable to regular tree languages.*

PROOF. Intuitively, we show that (i) LSTAs can accept sets of all perfect trees of an arbitrary height using their synchronization mechanism (while traditional TAs cannot) and, on the other hand, (ii) LSTAs cannot accept the set of all trees (with each branch having an arbitrary length), while traditional TAs can.

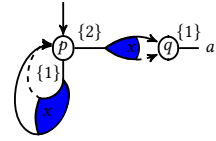

Fig. 18. LSTA recognizing a non-regular tree language

TA $\not\geq$ LSTA   Here we define an LSTA $\mathcal{A} = \langle \{p, q\}, \{x, a\}, \Delta, \{p\} \rangle$ accepting all perfect binary trees whose leaves are labeled $a$ and internal nodes are labeled $x$. The LSTA is defined in Fig. 18. We use the choices to force trees to make a consistent decision at each tree level: either all transitions generate leaves (choice 2) or internal nodes (choice 1). We will show that the language of this LSTA is not expressible by traditional TAs. We assume that there is a traditional TA $\mathcal{A}$ accepting all such states, which is an infinite set of perfect trees. We use $q_l^i$ and $q_r^i$ to denote the left and right bottom states, respectively, of the root transition of some of $\mathcal{A}$'s accepting run on a tree with height $i$. Since there are a finite number of states in $\mathcal{A}$, and there are infinitely many accepted trees of different heights, there must be two different heights, $i$ and $j$, such that $q_l^i = q_l^j$. This leads us to the conclusion that $q_l^i$ has (due to non-determinism) at least two possible subtrees with different heights below it. Since there is no synchronizing mechanism between states at the same level, $q_l^i$ can choose a subtree whose height is different from the height of $q_l^i$'s subtree. Therefore, $\mathcal{A}$ can accept a tree that is not perfect and, therefore, does not encode the set of perfect trees.

LSTA $\not\geq$ TA   For the other direction, consider the non-deterministic traditional TA $\mathcal{A} = \langle Q = \{q, q_a\}, \Sigma = \{x, a\}, \Delta, \mathcal{R} = \{q\} \rangle$ with transitions $q \rightarrow x(q, q)$, $q \rightarrow x(q_a, q_a)$, $q_a \rightarrow a$. Suppose that $\mathcal{L}(\mathcal{A})$ can be recognized by a LSTA $\mathcal{B} = \langle Q_{\mathcal{B}}, \Sigma, \Delta_{\mathcal{B}}, \mathcal{R}_{\mathcal{B}} \rangle$ with $|Q_{\mathcal{B}}| = m$ states. Let $n \gg 2m$. Consider the trees $T$ that are perfect above height $n$, with all nodes with height $\leq n$ labeled with $x$. Their subtrees below height $n$ are either of the form $t_x := x(x(a, a), x(a, a))$ or $t_a := x(a, a)$ (see for instance Fig. 19 for $n = 1$). At height $n$, there are $2^n$ tree nodes, and each can pick either subtree $t_a$ or $t_x$. Hence, there are $2^{2^n}$ such trees and these trees belong to $\mathcal{L}(\mathcal{A})$. By assumption, there are accepting runs of the LSTA $\mathcal{B}$ associated to these trees. Since each state of $\mathcal{B}$ at the top of a level of a tree run can only choose one unique transition, it follows that at the top of level $n$, $\mathcal{B}$ must have at least 2 states, say $q_x$ and $q_a$, in order to cover possible subtrees $t_x$ and $t_a$ respectively. Moreover, $q_x$ and $q_a$ can appear as left or right children of a transition used from level $n-1$ to level $n$ and there are $2^2$ possible pairs as bottoms of transitions. Thus, in order to cover all such trees, it requires at least $2^2$ states at the top of level $n-1$. In a similar fashion we can conclude that it requires at least $2^{2^m}$ states for level $n-m$ to cover all the tree runs for such trees. However, $2^{2^m} > m$ and it leads to a contradiction to the assumption that $|Q_{\mathcal{B}}| = m$. Thus, the theorem follows.                                              □

Table 1. Results of experiments. The columns **#q** and **#G** give the number of qubits and gates respectively of the circuit. For AutoQ, we give the times needed to obtain the LSTA representing all outputs of the circuit ($\mathrm{post}_C$), test the inclusion with the post-condition ($\subseteq$), and the sum of these times (total). The timeout was 5 min. We use colours to distinguish the best result and timeout/out-of-memory. "—" denotes that the tool was not applicable due to some limitation. The missing blocks for **FlipGate** are because the circuits did not contain CX gates. symQV timed out on all benchmarks from **Correct** and **FlipGate** and CaAL timed out on all benchmarks in **Correct** so we do not show their columns for those cases.

Column groups — **Correct**: AutoQ ($\mathrm{post}_C$, $\subseteq$, total), AutoQ-old, SliQSim, SV-Sim. **MissGate**: AutoQ ($\mathrm{post}_C$, $\subseteq$, total), AutoQ-old, symQV, CaAL. **FlipGate**: AutoQ ($\mathrm{post}_C$, $\subseteq$, total), AutoQ-old, CaAL.

| | #q | #G | postC | ⊆ | total | AutoQ-old | SliQSim | SV-Sim | postC | ⊆ | total | AutoQ-old | symQV | CaAL | postC | ⊆ | total | AutoQ-old | CaAL |
|---|---|---|---|---|---|---|---|---|---|---|---|---|---|---|---|---|---|---|---|
| BV-Sing | 96 | 241 | 0.2s | 0.0s | **0.2s** | 4.3s | 0.0s | OOM | 0.2s | 0.0s | **0.2s** | 4.8s | TO | TO | 0.2s | 0.0s | **0.2s** | 4.4s | TO |
| | 97 | 243 | 0.2s | 0.0s | **0.2s** | 4.6s | 0.0s | OOM | 0.2s | 0.0s | **0.2s** | 4.8s | TO | TO | 0.2s | 0.0s | **0.2s** | 4.6s | TO |
| | 98 | 246 | 0.3s | 0.0s | **0.3s** | 4.7s | 0.0s | OOM | 0.2s | 0.0s | **0.2s** | 5s | TO | TO | 0.3s | 0.0s | **0.3s** | 4.7s | TO |
| | 99 | 248 | 0.2s | 0.0s | **0.2s** | 4.9s | 0.0s | OOM | 0.3s | 0.0s | **0.3s** | 5.1s | TO | TO | 0.3s | 0.0s | **0.3s** | 4.8s | TO |
| | 100 | 251 | 0.3s | 0.0s | **0.3s** | 5s | 0.0s | OOM | 0.2s | 0.0s | **0.2s** | 5.2s | TO | TO | 0.2s | 0.0s | **0.2s** | 5s | TO |
| BV-All | 19 | 30 | 0.0s | 0.0s | **0.0s** | 2.8s | TO | TO | 0.0s | 0.0s | **0.0s** | 2.9s | TO | TO | | | | | |
| | 21 | 33 | 0.0s | 0.0s | **0.0s** | 8s | TO | TO | 0.0s | 0.0s | **0.0s** | 7.3s | TO | TO | | | | | |
| | 23 | 36 | 0.0s | 0.0s | **0.0s** | 23s | TO | TO | 0.0s | 0.0s | **0.0s** | 44s | TO | TO | | | | | |
| | 25 | 39 | 0.0s | 0.0s | **0.0s** | 1m11s | TO | TO | 0.0s | 0.0s | **0.0s** | 1m58s | TO | TO | | | | | |
| | 27 | 42 | 0.0s | 0.0s | **0.0s** | 4m45s | TO | TO | 0.0s | 0.0s | **0.0s** | TO | TO | TO | | | | | |
| GHZ-Sing | 64 | 64 | 0.0s | 0.0s | **0.0s** | 0.3s | 0.0s | OOM | 0.1s | 0.0s | **0.1s** | 0.4s | TO | TO | 0.1s | 0.0s | **0.1s** | 0.3s | TO |
| | 128 | 128 | 0.1s | 0.0s | **0.1s** | 2.5s | 0.0s | OOM | 0.1s | 0.0s | **0.1s** | 2.5s | TO | TO | 0.1s | 0.0s | **0.1s** | 2.7s | TO |
| | 256 | 256 | 0.8s | 0.0s | **0.8s** | 20s | 0.0s | OOM | 0.8s | 0.0s | **0.8s** | 20s | TO | TO | 0.7s | 0.0s | **0.7s** | 20s | TO |
| | 384 | 384 | 1.9s | 0.0s | **1.9s** | 1m10s | 0.1s | OOM | 1.9s | 0.0s | **1.9s** | 1m12s | TO | TO | 1.7s | 0.0s | **1.7s** | 1m9s | TO |
| | 512 | 512 | 3.6s | 0.0s | **3.6s** | 2m49s | 0.1s | OOM | 3.6s | 0.0s | **3.6s** | 2m59s | TO | TO | 3.3s | 0.0s | **3.3s** | 2m52s | TO |
| GHZ-All | 8 | 8 | 0.0s | 0.0s | **0.0s** | 0.1s | 1.5s | OOM | 0.0s | 0.0s | **0.0s** | 0.2s | TO | 30s | 0.0s | 0.0s | **0.0s** | 0.1s | 32s |
| | 16 | 16 | 0.0s | 0.0s | **0.0s** | TO | TO | 1m37s | 0.0s | 0.0s | **0.0s** | TO | TO | TO | 0.0s | 0.0s | **0.0s** | TO | 55s |
| | 32 | 32 | 0.0s | 0.0s | **0.0s** | TO | TO | TO | 0.0s | 0.0s | **0.0s** | TO | TO | 1m28s | 0.0s | 0.0s | **0.0s** | TO | 1m27s |
| | 64 | 64 | 0.1s | 0.0s | **0.1s** | TO | TO | TO | 0.1s | 0.0s | **0.1s** | TO | TO | TO | 0.1s | 0.0s | **0.1s** | TO | 3m7s |
| | 128 | 128 | 0.6s | 0.0s | **0.6s** | TO | TO | OOM | 0.5s | 0.0s | **0.5s** | TO | TO | TO | 0.6s | 0.0s | **0.6s** | TO | TO |
| Grover-Sing | 24 | 5215 | 1.3s | 0.0s | **1.3s** | 10s | TO | TO | 1.3s | 0.0s | **1.3s** | 10s | TO | — | 1.5s | 0.0s | **1.5s** | 11s | — |
| | 28 | 12217 | 3.8s | 0.0s | **3.8s** | 32s | TO | TO | 3.9s | 0.0s | **3.9s** | 32s | TO | — | 4.1s | 0.0s | **4.1s** | 35s | — |
| | 32 | 28159 | 10s | 0.0s | **10s** | 1m28s | TO | OOM | 16s | 0.0s | **16s** | 1m28s | TO | — | 11s | 0.0s | **11s** | 1m43s | — |
| | 36 | 63537 | 26s | 0.0s | **26s** | 4m3s | TO | OOM | 45s | 0.0s | **45s** | TO | TO | — | 28s | 0.0s | **28s** | 4m36s | — |
| | 40 | 141527 | 1m8s | 0.0s | **1m8s** | TO | TO | OOM | 1m28s | 0.0s | **1m28s** | TO | TO | — | 1m13s | 0.0s | **1m13s** | TO | — |
| Grover-All | 18 | 357 | 0.2s | 0.0s | **0.2s** | 3.1s | TO | 38s | 0.2s | 0.0s | **0.2s** | 3.6s | TO | TO | 0.5s | 0.0s | **0.5s** | 10s | TO |
| | 21 | 552 | 0.8s | 0.0s | **0.8s** | 10s | TO | TO | 0.9s | 0.0s | **0.9s** | 12s | TO | — | 1.8s | 0.0s | **1.8s** | 31s | — |
| | 24 | 939 | 3.0s | 0.1s | **3.1s** | 38.1s | TO | TO | 3.5s | 0.1s | **3.1s** | 44s | TO | — | 7.1s | 0.2s | **7.3s** | TO | — |
| | 27 | 1492 | 9.7s | 0.8s | **10.5s** | 2m20s | TO | TO | 9.8s | 0.7s | **10.5s** | 3m38s | TO | — | 25s | 1.1s | **26.1s** | TO | — |
| | 30 | 2433 | 36s | 3.3s | **39.3s** | TO | TO | OOM | 36s | 3.2s | **39.2s** | TO | TO | — | 1m28s | 4.2s | **1m32s** | TO | — |
| H2 | 12 | 24 | 0.0s | 0.0s | **0.0s** | 40s | 1m51s | 0.5s | 0.0s | 0.0s | **0.0s** | 39s | TO | TO | | | | | |
| | 13 | 26 | 0.0s | 0.0s | **0.0s** | 2m35s | TO | 2.1s | 0.0s | 0.0s | **0.0s** | 2m15s | TO | OOM | | | | | |
| | 64 | 128 | 0.2s | 0.0s | **0.2s** | TO | TO | OOM | 0.2s | 0.0s | **0.2s** | TO | TO | TO | | | | | |
| | 128 | 256 | 0.9s | 0.0s | **0.9s** | TO | TO | OOM | 1.0s | 0.0s | **1s** | TO | TO | TO | | | | | |
| | 256 | 512 | 4.4s | 0.0s | **4.4s** | TO | TO | OOM | 4.9s | 0.0s | **4.9s** | TO | TO | TO | | | | | |
| HXH | 10 | 30 | 0.0s | 0.0s | **0.0s** | 1.8s | 11s | 0.0s | 0.0s | 0.0s | **0.0s** | 1.1s | TO | TO | | | | | |
| | 11 | 33 | 0.0s | 0.0s | **0.0s** | 4.3s | 34s | 0.2s | 0.0s | 0.0s | **0.0s** | 5.3s | TO | TO | | | | | |
| | 12 | 36 | 0.0s | 0.0s | **0.0s** | 11.1s | 1m51s | 0.7s | 0.0s | 0.0s | **0.0s** | 15s | TO | TO | | | | | |
| | 13 | 39 | 0.0s | 0.0s | **0.0s** | 36.3s | TO | 2.9s | 0.0s | 0.0s | **0.0s** | 1m51s | TO | TO | | | | | |
| | 99 | 297 | 0.7s | 0.0s | **0.7s** | TO | TO | TO | 0.7s | 0.0s | **0.7s** | TO | TO | TO | | | | | |
| MCToffoli | 16 | 15 | 0s/0s | 0s/0s | **0.0s/0.0s** | 0.0s/0.0s | TO | 7.7s | 0s/0s | 0s/0s | **0.0s/0.0s** | 0.0s/0.0s | TO | TO | 0s/0s | 0s/0s | **0.0s/0.0s** | 0.0s/0.0s | TO |
| | 20 | 19 | 0s/0s | 0s/0s | **0.0s/0.0s** | 0.0s/0.0s | TO | TO | 0s/0s | 0s/0s | **0.0s/0.0s** | 0.0s/0.0s | TO | TO | 0s/0s | 0s/0s | **0.0s/0.0s** | 0.0s/0.0s | TO |
| | 24 | 23 | 0s/0s | 0s/0s | **0.0s/0.0s** | 0.0s/0.0s | TO | TO | 0s/0s | 0s/0s | **0.0s/0.0s** | 0.0s/0.0s | TO | TO | 0s/0s | 0s/0s | **0.0s/0.0s** | 0.0s/0.0s | TO |
| | 28 | 27 | 0s/0s | 0s/0s | **0.0s/0.0s** | 0.1s/0.1s | TO | TO | 0s/0s | 0s/0s | **0.0s/0.0s** | 0.1s/0.1s | TO | TO | 0s/0s | 0s/0s | **0.0s/0.0s** | 0.1s/0.1s | TO |
| | 32 | 31 | 0s/0s | 0s/0s | **0.0s/0.0s** | 0.1s/0.2s | TO | TO | 0s/0s | 0s/0s | **0.0s/0.0s** | 0.1s/0.1s | TO | TO | 0s/0s | 0s/0s | **0.0s/0.0s** | 0.2s/0.2s | TO |
| Grover-Iter | 3 | 36 | 0.0s | 0.0s | **0.0s** | 0.3s | — | — | 0.0s | 0.1s | **0.1s** | TO | 4m7s | — | | | | | |
| | 32 | 157 | 0.0s | 0.0s | **0.0s** | 2.0s | — | — | 0.0s | 0.1s | **0.1s** | 2.3s | TO | — | | | | | |
| | 100 | 445 | 2.2s | 0.0s | **2.2s** | 49s | — | — | 2.4s | 0.1s | **2.5s** | 55s | TO | — | | | | | |
| | 150 | 671 | 7.4s | 0.1s | **7.5s** | 3m35s | — | — | 7.7s | 0.1s | **7.8s** | TO | TO | — | | | | | |
| | 200 | 895 | 17s | 0.1s | **17.1s** | TO | | | 18s | 0.1s | **18.1s** | TO | TO | — | | | | | |

## 7 EXPERIMENTAL EVALUATION

We implemented our LSTA-based verification framework as an updated version of AutoQ[6]. It is written in C++ and combines the three needed components: the LSTA symbolic representation from Sections 3 and 4, the gate operations from Section 5, and the entailment checking algorithm from Section 6. As an input, AutoQ takes a quantum circuit $C$ in the OpenQASM format, a pre-condition LSTA P, and a post-condition LSTA Q. Starting from the pre-condition LSTA P, AutoQ reads the quantum gates from $C$ one by one and executes them symbolically to obtain an output LSTA R. AutoQ then checks if $\mathcal{L}(\mathsf{R}) \subseteq \mathcal{L}(\mathsf{Q})$ using the entailment testing algorithm from Section 6. When the test fails, AutoQ reports a reachable quantum state violating the post-condition for diagnostics. We use a precise complex number representation similar to those in [18, 50, 63], with an extension to allow a wider range of angles. For rotation gates such as $R_X(\theta)$ and $R_Z(\theta)$, we allow $\theta$ in the form of $\frac{n}{4}\pi$ for $n \in \mathbb{Z}$.

[6] https://github.com/fmlab-iis/AutoQ

*Tools.* We compared the new AutoQ with several state-of-the-art tools. Among these, the only tool directly comparable (as it also performs automated Hoare-style verification) to AutoQ is its predecessor [17, 18], which uses an approach based on traditional tree automata. Second, we compared AutoQ against two symbolic quantum circuit verification tools: symQV [7], which is based on the SMT theory of reals, and CaAL [19], which is based on an extended SMT theory of arrays. In our evaluation, we specified equivalent functional correctness properties for each benchmark example in their respective specification languages. Third, when the precondition involves a finite number of quantum states, we can solve the verification task by using *quantum circuit simulators* to simulate all allowed initial quantum states and validate against the postcondition. Therefore, we also compared AutoQ with the state vector-based simulator SV-Sim [37] and the decision diagram-based simulator SliQSim [50]. Validation against postcondition is, however, hard for these two simulators because printing the final state as an explicit vector takes too much time. Therefore, we report the time required for circuit simulation and omit the time for validation against the postcondition. This gives us a conservative under-approximation of the time the simulators would need for verification.

*Benchmarks.* We compared all tools on the following set of benchmarks:

- **BV-Sing and BV-All:** Bernstein-Vazirani's algorithm [8] (Section 4.1) with one hidden string and with all possible hidden strings of length $n$, respectively. For BV-Sing, the hidden string is in the form of $|1010\ldots\rangle$; for BV-All, hidden strings are decided by input.
- **GHZ-Sing and GHZ-All:** The GHZ circuit [28] with the pure zero input state $|0^n\rangle$ and with all possible input states $\{|s_1 0 s_2 0 \ldots s_n 01\rangle \mid s_1, s_2, \ldots s_n \in \mathbb{B}\}$.
- **Grover-Sing and Grover-All:** Grover's search [29] circuit for a single oracle and for all possible oracles of length $n$. For the former, the hidden item is $|0101\ldots\rangle$. We verify that the output amplitudes match the expected values.
- **Grover-Iter:** Verification of one iteration of Grover's search w.r.t. the symbolic property that the amplitude of the secret is amplified (Section 4.4).
- **H2 and HXH:** the former has two consecutive H gates for each qubit in an $n$-qubit quantum circuit and the latter has one additional X gate in between the two H gates. Both use all possible computational basis states as the pre-condition.
- **MCToffoli:** circuits implementing multi-control Toffoli gates of size $n$ using a variation of Nielsen and Chuang's decomposition [44] with standard Toffoli gates. The pre- and post-conditions are those from Section 4.2. We always give a pair of results: for $k=0/k=1$.

We further consider the following three scenarios:

- **Correct:** verification of a correct circuit.
- **MissGate:** finding a bug in a circuit obtained from a correct one by removing a random gate.
- **FlipGate:** finding a bug in a circuit obtained from a correct circuit by selecting a random CX gate and swapping its control and target qubits.

*Evaluation and results.* We conducted all our experiments on a server running Ubuntu 22.04.3 LTS with an AMD EPYC 7742 64-core processor (1.5 GHz), 2 TiB of RAM, and a 1 TB SSD; the timeout was 5 min. We give the results in Table 1.

In the CORRECT scenario, we can see that with the exception of the BV-SING and GHZ-SING benchmarks (which are much easier because the precondition is a singleton set), AUTOQ outperforms other tools by several orders of magnitude and scales much better with increasing sizes of the circuits. Even on the two mentioned benchmarks, AUTOQ is still competitive; although slower than SLIQSIM, it still performs much better than SV-SIM (which fails due to running out of memory). Comparing AUTOQ with the closest competitor, AUTOQ-OLD, which is based on traditional tree automata, we can see that the use of LSTAs in most cases yields a dramatic speed-up, affirming the usefulness of LSTAs. Neither SYMQV nor CAAL managed to finish on any of the benchmarks here.

In the MISSGATE and FLIPGATE scenarios, AUTOQ again outperforms AUTOQ-OLD by several orders of magnitude. Contrary to the T scenario, SYMQV and CAAL manage to finish on a few benchmarks (SYMQV: 1 benchmark in MISSGATE; CAAL: 3 benchmarks in MISSGATE and 4 benchmarks in FLIPGATE), successfully catching the bugs. The performance of AUTOQ is, however, still much better.

We also demonstrate the expressiveness of LSTA as a specification language and the versatility of AUTOQ by using it to check the equivalence of circuits. AUTOQ is not optimized for this task, and as expected, it is not as efficient as specialized circuit equivalence checkers such as SLIQEC [16, 55] and QCEC [13], but it can still handle several practical cases. The largest case (in the number of qubits) we can handle is the circuit add64_164 from REVLIB [56], with 193 qubits and 256 gates before optimization. We used QISKIT [5] to transpile and optimize it and then check equivalence of the optimized circuit and the original one (cf. Section 4.3). AUTOQ was able to verify the equivalence between the two versions within 24 seconds.

We emphasize that the LSTA-based approach is the most efficient when the number of distinct amplitude values of reachable quantum states is small. For circuits such as the quantum Fourier transform (QFT), where the reachable quantum states have an exponential number of distinct amplitude values, our approach with the current state encoding struggles. A potential alternative encoding is discussed in Section 8.3.

## 8   TOWARDS PARAMETERIZED VERIFICATION OF QUANTUM CIRCUITS

Let us now move to verification of *parameterized circuits*, i.e., checking correctness of a (typically) infinite family of quantum circuits with a similar structure (differing in the number of qubits). To the best of our knowledge, parameterized verification of quantum circuits has not been addressed by any fully automated approach so far. In this section, we will show that the use of LSTAs enables automated verification of a certain class of parameterized circuits. In particular, we will demonstrate how to use LSTAs to verify the parameterized GHZ circuit [28] and proceed to the verification of other circuits used in practice, such as circuits implementing parameterized *fermionic unitary evolution*. Our modelling and experiments with yet another class of parameterized circuits performing *diagonal Hamiltonian simulation* can be found in Appendix A.3 of the supplementary material.  At the end, we will discuss the challenges and a direction towards a complete framework for parameterized circuit verification.

### 8.1   Verification of the Parameterized GHZ Circuit

An $n$-qubit *GHZ state* (for $n \geq 1$) is a state of the form $\frac{|0^n\rangle+|1^n\rangle}{\sqrt{2}}$ and an LSTA recognizing such states can be found in Fig. 20a. An $n$-qubit *GHZ circuit* is shown in Fig. 20b; it first executes the $H_1$ gate and then the sequence of gates $CX_2^1, CX_3^2, \ldots, CX_n^{n-1}$. The number of CX gates involved is $n-1$. The circuit transforms a quantum state of the form $|0^n\rangle$ (represented by the LSTA in Fig. 6) to $\frac{|0^n\rangle+|1^n\rangle}{\sqrt{2}}$ (represented by the LSTA in Fig. 20a). To execute this family of circuits, we need to

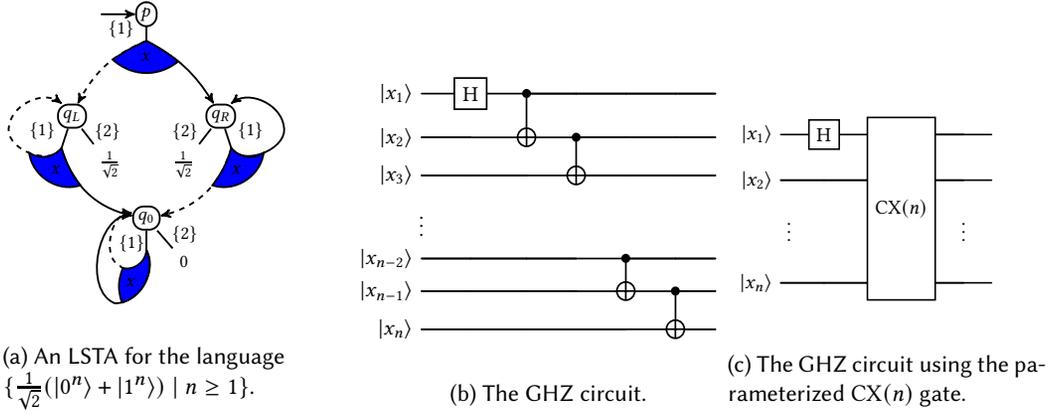

(a) An LSTA for the language
$\{\frac{1}{\sqrt{2}}(|0^n\rangle + |1^n\rangle) \mid n \geq 1\}$.

(b) The GHZ circuit.

(c) The GHZ circuit using the parameterized CX($n$) gate.

Fig. 20. An LSTA for the set of GHZ states (a), the GHZ circuit (b), and the parameterized CX($n$) gate (c). We note that in our uses of the CX($n$) gate (and other parameterized gates), the inputs are given by the order of qubits (i.e., corresponding to the schema in (b)).

support parameterized quantum gates, which is generally hard but still feasible in our framework for some cases. We first introduce the *parameterized* CX($n$) *gate*, which implements the sequence of $n - 1$ CX gates on $n$ qubits (cf. Figs. 20b and 20c). We will now describe how the effect of such a parameterized gate is computed on an LSTA representing a set of quantum states.

CX($n$) gate   In Algorithm 6, we introduce a procedure to build an LSTA CX($n$)($\mathcal{A}$) representing the set of states after executing the CX($n$) gate on all qubits. Recall that the effect of applying the CX$_{i+1}^i$ gate on a state has the same effect of applying an $X^{i+1}$ on the 1-subtree below every node labeled $x_i$. Moreover, performing $X$ gate to all qubits of a state, i.e., the $X^{\otimes n}$ gate, reverses the corresponding vector of amplitudes, e.g., from $(a, b, c, d)^T$ to $(d, c, b, a)^T$ or, equivalently, reverses all amplitude values of the tree. Thus, in Algorithm 6, we first construct a primed version $\overline{\mathcal{A}} = X^{\otimes n}(\mathcal{A})$, which encodes trees obtained by reversing the amplitude values of states in $\mathcal{L}(\mathcal{A})$. Such an LSTA $\overline{\mathcal{A}}$ can be constructed from $\mathcal{A}$ by swapping the two bottom states of all non-leaf transitions (Lines 1–2). The resulting automaton is then obtained by combining $\mathcal{A}$ and $\overline{\mathcal{A}}$ such that the right-hand bottom state of $\mathcal{A}$ is reconnected to the corresponding state of $\overline{\mathcal{A}}$ (Line 3), i.e., jumping to $\overline{\mathcal{A}}$ when the control qubit has value 1, and the left-hand bottom state of $\overline{\mathcal{A}}$ is reconnected to the corresponding state of $\mathcal{A}$ (Line 4), i.e., jumping back to the original LSTA when control qubit has value 0. Here, $\Delta_0$ and $\overline{\Delta_0}$ are the sets of leaf transitions.

THEOREM 18. $\mathcal{L}(\text{CX}(n)(\mathcal{A})) = \{\text{CX}_n^{n-1}(\cdots \text{CX}_2^1(T) \cdots) \mid T \in \mathcal{L}(\mathcal{A})\}$ *and* $|\text{CX}(n)(\mathcal{A})| = 2|\mathcal{A}|$.

To verify the parameterized GHZ circuit, we begin by using the precondition $P$ provided in the LSTA of Fig. 6. Next, we apply the H$_1$ operation using Algorithm 1 and reduce the output. We then run Algorithm 6 on this LSTA to obtain $\text{post}_C(P)$. Finally, we check for inclusion against the postcondition $Q$ in Fig. 20a.

Non-parameterized gates   Our construction of (non-parameterized) LSTA quantum gate operations from Section 5 targeting qubit $x_t$ requires explicitly assigned index numbers to the first $t$ qubits. We can obtain the explicit qubit numbering by *unfolding* the transitions for $t$ layers from the root states and labeling them in order from $x_1$ to $x_t$.[7] We can unfold the last qubit $x_t$ using a

---

[7]One way to do this is to create $t$ copies of the LSTA $\mathcal{A}$ and name them $\mathcal{A}_1$ to $\mathcal{A}_t$. Next, we rename the transition symbols of $\mathcal{A}_i$ from $x$ to $x_i$ and states from $s$ to $s_i$ for $1 \leq i \leq t$. After that, we connect these LSTAs and form an unfolded one.

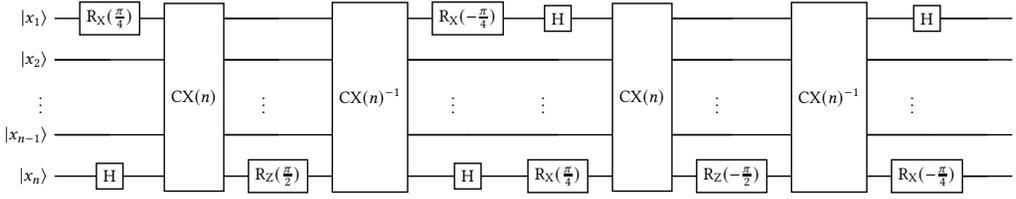

(a) A standard circuit performing a single fermionic excitation.

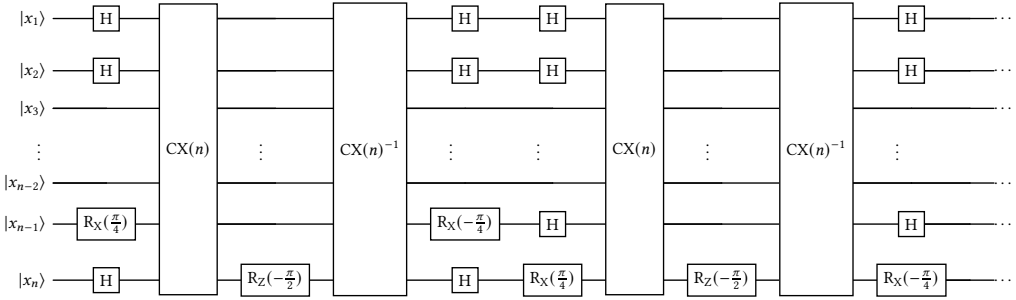

(b) Part of the circuit performing a double fermionic excitation. See [60, Fig. 11] for the whole circuit.

Fig. 21. Circuits performing single and double fermionic excitations. The order of qubits for the CX($n$) and CX($n$)$^{-1}$ gates correspond to the order of qubits in Figs. 20b and 22.

similar construction. After applied a quantum gate, the LSTA can be *folded* again by replacing all indexed qubit symbols $x_i$ with $x$ and running a reduction algorithm.

## 8.2  Verification of Fermionic Unitary Evolutions

Two major challenges in the practical realization of the *variational quantum eigensolver* on *noisy intermediate-scale quantum* (NISQ) computers are the design of *ansatz* states (i.e., states that serve as the initial *guess* of the final value) and the construction of efficient circuits to prepare these states for a parameterized number of qubits. Most ansatz states considered by the quantum chemistry and material science communities correspond to

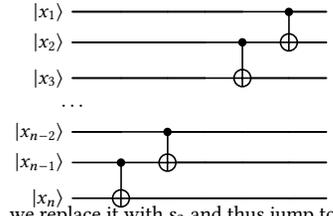

Starting from $\mathcal{A}_1$, for each state $s_1$ that can be reached from root states in one step, we replace it with $s_2$ and thus jump to $\mathcal{A}_2$. We then find a one-step reachable state from $s_2$ and jump to $s_3$, and repeat this process until we jump from $s_3$ to $\mathcal{A}$. Figs 22. The CX($n$) gate.

---

**Algorithm 6:** Application of the CX($n$) gate on an LSTA

---

**Input:** An LSTA $\mathcal{A} = \langle Q, \Sigma, \Delta, \mathcal{R} \rangle$ representing a parameterized set of quantum states
**Output:** An LSTA CX($n$)($\mathcal{A}$)

1  Build $\overline{\mathcal{A}} = \langle \overline{Q}, \Sigma, \overline{\Delta}, \overline{\mathcal{R}} \rangle$, a barred copy of $\mathcal{A}$ with

2  $\overline{\Delta} = \{\overline{q} \text{ -}C\text{→ } f(\overline{q_r}, \overline{q_l}) \mid q \text{ -}C\text{→ } f(q_l, q_r) \in \Delta\} \cup \{\overline{q} \text{ -}C\text{→ } c \mid q \text{ -}C\text{→ } c \in \Delta\}; \text{ // } \overline{\mathcal{A}} = X^{\otimes n}(\mathcal{A})$

3  $\Delta^R := \{q \text{ -}C\text{→ } f(q_l, \overline{q_r}) \mid q \text{ -}C\text{→ } f(q_l, q_r) \in \Delta\} \cup$

4  $\{\overline{q} \text{ -}C\text{→ } f(q_r, \overline{q_l}) \mid \overline{q} \text{ -}C\text{→ } f(\overline{q_r}, \overline{q_l}) \in \overline{\Delta}\} \cup \Delta_0 \cup \overline{\Delta_0};$

5  **return** $\mathcal{A}^R = \langle Q \uplus \overline{Q}, \Sigma, \Delta^R, \mathcal{R} \rangle;$

---

---

**Algorithm 7:** Application of the $CX(n)^{-1}$ gate on an LSTA

---

**Input:** An LSTA $\mathcal{A} = \langle Q, \Sigma, \Delta, \mathcal{R} \rangle$ representing a parameterized set of quantum states
**Output:** An LSTA $CX(n)^{-1}(\mathcal{A})$

1  Build $\overline{\mathcal{A}} = \langle \overline{Q}, \Sigma, \overline{\Delta}, \overline{\mathcal{R}} \rangle$ with

2  $\overline{\Delta} = \{\overline{q} \, \text{-}C\text{→} f(\overline{q_r}, \overline{q_l}) \mid q \, \text{-}C\text{→} f(q_l, q_r) \in \Delta\} \cup \{\overline{q} \, \text{-}C\text{→} c \mid q \, \text{-}C\text{→} c \in \Delta\};$ // $\overline{\mathcal{A}} = X^{\otimes n}(\mathcal{A})$

3  $\Delta^R := \{q \, \text{-}C\text{→} f(q_l, \overline{q_r}) \mid q \, \text{-}C\text{→} f(q_l, q_r) \in \Delta\} \cup$

4      $\{\overline{q} \, \text{-}C\text{→} f(\overline{q_r}, q_l) \mid \overline{q} \, \text{-}C\text{→} f(\overline{q_r}, \overline{q_l}) \in \overline{\Delta}\} \cup \Delta_0 \cup \overline{\Delta}_0;$

5  **return** $\mathcal{A}^R = \langle Q \uplus \overline{Q}, \Sigma, \Delta^R, \mathcal{R} \rangle;$

---

applying a series of *fermionic unitary evolutions* to an initial reference state [60]. These evolutions are referred as single and double *fermionic excitations* and can be transformed via *Jordan-Wigner* encoding into the quantum circuits in Fig. 21a and Fig. 21b [60]. Note that the circuits use the parameterized $CX(n)^{-1}$ gate, which is given in Fig. 22 and whose implementation in our framework is given in Algorithm 7. Both circuits can be verified in our framework against the precondition $\{|0^n\rangle \mid n \geq 1\}$ (Fig. 6).

$\boxed{CX(n)^{-1} \text{ gate}}$ Algorithm 7 is very similar to Algorithm 6. The only difference between is that on Line 4, the right-hand bottom state of $\overline{\mathcal{A}}$ is reconnected to the corresponding state of $\mathcal{A}$, not the left-hand side state, i.e., we jump back to the original LSTA when the value of the control qubit is 1.

THEOREM 19. $\mathcal{L}(CX(n)^{-1}(\mathcal{A})) = \{CX_2^1(\cdots CX_n^{n-1}(T)\cdots) \mid T \in \mathcal{L}(\mathcal{A})\}$ *and* $|CX(n)^{-1}(\mathcal{A})| = 2|\mathcal{A}|.$

$\boxed{\text{Other parameterized gates}}$ Our framework supports not only $CX(n)$ and $CX(n)^{-1}$ gates, but also $X^{\otimes n}$, $(D^{1,\omega_N^m})^{\otimes n}$, and some alternating patterns of CX gates. More details can be found in Appendix A.4.

## 8.3 Evaluation and Discussion

We extended AUTOQ to support parameterized verification and used it to verify the circuits from Figs. 20c, 21a and 21b, and a circuit for diagonal Hamiltonian simulation (DHS; cf. Appendix A.3). The results are given in Table 2. All of these families of circuits could be verified efficiently. Even though the sets of encoded quantum states are infinitely large, their symbolic representation is compact (less than 10 states and 20 transitions).

Our current approach supports a limited set of parameterized quantum gates. Specifically, the gate $H^{\otimes n}$ is currently beyond our capability because applying it to the states $|0^n\rangle$ would create an infinite set of constant symbols $\{\frac{1}{\sqrt{2}^k} \mid k \in \mathbb{N}\}$. To overcome this challenge, we propose introducing a *state variable s* to represent the current basis state[8] (i.e., the values of qubits on the branch in the tree) and using it in the leaf symbol. E.g., this would allow us to use the singleton set $\{\frac{1}{\sqrt{2}^{|s|}}\}$ as the leaf symbols, where $|s|$ represents the number of qubits. It also makes the leaf symbols after executing a *quantum Fourier transform* (QFT) circuit more concise. In the case of two qubits, QFT converts a quantum state $(a, b, c, d)^T$ to a tree with all leaves labeled with $(a + b\omega^{1\cdot s} + c\omega^{2\cdot s} + d\omega^{3\cdot s})$, which can be made concise with the proposed encoding. We are quite enthusiastic about exploring this direction further in the future, as it promises to extend our approach's capabilities and open new possibilities for parameterized quantum circuit verification.

Table 2. Results for the verification of parameterized circuits

| | #G | post$_C$ | $\subseteq$ | total |
|---|---|---|---|---|
| | | AUTOQ | | |
| Fig. 20c | 2 | 0.0s | 0.0s | **0.0s** |
| DHS | 4 | 0.0s | 0.0s | **0.0s** |
| Fig. 21a | 14 | 0.3s | 0.0s | **0.3s** |
| Fig. 21b | 88 | 1m54s | 0.0s | **1m54s** |

---

[8] A similar concept of using state variables is utilized in the path-sum approach [4, 15].

## 9 RELATED WORK

We split the related works into two parts. First, we provide an overview of approaches of quantum circuits analysis and explain where our approach stands among existing tools. Then we will provide a detailed comparison of the quantum predicates we use and those of D'Hondt and Panangaden [23], which are used in the standard quantum Hoare logic of Ying [59].

### 9.1 Approaches for Quantum Circuit Analysis

In recent years, many techniques for analyzing, simulating, and verifying quantum circuits and programs have emerged in the formal methods community.

*Symbolic Quantum Circuit Verifiers:* As mentioned in the introduction, these tools are fully automated, flexible in specifying verification properties, and provide precise bug traces. The closest work to ours is [17, 18], which uses traditional tree automata [22] to encode predicates representing sets of quantum states. We use the same verification framework as in [17, 18], employing automata to specify pre- and post-conditions and developing algorithms for the symbolic execution of quantum gates. LSTAs solve the scalability issues in many cases where the size of the representation using traditional TAs blew up (cf. Section 7), due to the introduction of level synchronization. Moreover, our single qubit gate operations are quadratic, while those of [17, 18] are exponential. There are two more tools that belong to this category: SYMQV [7] is based on *symbolic execution* [34] to verify input-output relationship with queries discharged by SMT solvers over the theory of reals. The SMT array theory approach [19] improved the previous by allowing a polynomial-sized circuit encoding but still faces a similar scalability problem. As we can see from our experiments, AUTOQ significantly outperforms all of the above tools in this category by several orders of magnitude. Another scalable fully automated approach for analysis of quantum circuits is *quantum abstract interpretation* [46, 61]. Our approach is more suitable for bug hunting as it does not rely on overapproximation, which is used by quantum abstract interpretation.

*Deductive quantum circuit/program verifiers:* These tools allow the verification of quantum programs w.r.t. very expressive specification languages (to be discussed in the next part). One most prominent family of approaches is based on the so-called *quantum Hoare logic* [25, 38, 51, 59, 62]. These approaches, however, require significant manual work and user expertise (they are often based on the use of interactive theorem provers such as ISABELLE [45] and COQ [9]). The approach implemented in QBRICKS [15] tries to alleviate this issue by generating proof obligations and discharging them automatically using SMT solvers, but this still requires a lot of interaction with the user (e.g., when verifying Grover's search for an arbitrary number of qubits, the experiment needed over one hundred user interactions).

*Quantum circuit equivalence checkers:* These tools are usually fully automated but are limited to only checking equivalence of two circuits. In contrast, our approach is flexible in specifying custom properties. Equivalence checkers are based on several approaches. One approach is based on the *ZX-calculus* [21], which is a graphical language used for reasoning about quantum circuits. The *path-sum* approach (implemented, e.g., within the tool FEYNMAN [4]) uses rewrite rules. Pre-computed equivalence sets are used to prove equivalence in QUARTZ [58]. QCEC [13] is an equivalence checker that uses decision diagrams and ZX-calculus, and SLIQEC [16, 55] also uses decision diagrams for (partial) equivalence checking. An approach based on working with the so-called *stabilizer states* [49] can be used to verify the equivalence of circuits with Clifford gates in polynomial time.

*Quantum circuit simulators:* Simulators can be categorized into decision diagram-based [41, 48, 50, 63], vector-based [37], and tensor-network-based [39, 53]. These simulators are designed to compute the circuit output for a single input state. They can also be used to verify the behavior of quantum circuits for a finite number of input states by simulating all. Our experiments, however, show that when preconditions permit exponentially many states, these simulators do not scale well to large numbers of qubits. Our approach can be viewed as a generalization of the decision diagram-based method, allowing us to handle sets of input states simultaneously. While tensor networks (TN) have proven highly efficient in simulations for physics applications and "quantum supremacy" benchmarks, they have yet to be adapted for verification purposes. Investigating TN-based methods for verification presents a promising avenue for future research. In particular, exploring how TN simulation techniques, such as the decision diagram variant [31], could be extended to automata for verification is a compelling direction for future work.

### 9.2 Comparison with the Quantum Predicates of D'Hondt and Panangaden [23]

In the set-based approach ([17, 18] and this paper), a predicate maps a *pure state* to 0 or 1. Automata are used as compact representations of set-based predicates. A tree (representing a pure state) is accepted by an automaton iff the predicate maps the corresponding pure state to 1. On the other hand, in [23] and the quantum Hoare logic of Ying [59], a predicate is a mapping from a *mixed state* (a distribution over pure states) to a real value between 0 and 1, capturing the likelihood of mixed states satisfying the specified condition. An important advantage of the set-based approach is that it includes an efficient automated verification algorithm, whereas standard quantum Hoare logic typically relies on an interactive deductive method, requiring more manual effort. Furthermore, we demonstrate the potential of the set-based approach for automatic parameterized verification of quantum circuits.

The two type of predicates are incomparable in terms of expressive power. The predicates used in [23] are "continuous" in the sense that two "similar" mixed states (say, up to trace distance) will be mapped to two similar values. In contrast, our predicates are "discrete:" For two "similar" pure states, our predicates can map one to 1 and the other to 0. There are pros and cons of both systems. For instance, the "continuous" predicates [23, 59] can do a more fine-grained analysis of *quantum programs*[9]; they can talk about the probability of a pure state at the output. They can also talk about the expected value of a certain observation. In contrast, the "discrete" predicate can only verify state non-zero reachability, but not the precise probability. Conversely, continuous predicates cannot precisely describe discrete sets of states or vectors due to their continuous nature. Discrete predicates can specify the identity of a circuit (and thus circuit equivalence) using a set of $2^n$ linearly independent vectors simultaneously as the pre- and post-conditions (cf. Section 4), encoded with a linear number of transitions relative to $n$. It might seem natural to represent a set of pure states $\{|\phi_i\rangle \mid 0 \le i \le n\}$ by summing their outer products, $\sum_{0 \le i \le n} |\phi_i\rangle \langle \phi_i|$, and use this as a predicate in Ying's quantum Hoare logic [59]. However, doing so can lead to ambiguity, as different sets of pure states may correspond to the same predicate. For instance, the sets $\{|0\rangle, |1\rangle\}$ and $\{|-\rangle, |+\rangle\}$[10] yield the same outer product sum, failing to distinguish between distinct pure states.

---

[9]Quantum programs are an extension of quantum circuits that allows branching and loop statements.
[10]$|-\rangle = \frac{1}{\sqrt{2}}(|0\rangle + |1\rangle)$ and $|+\rangle = \frac{1}{\sqrt{2}}(|0\rangle - |1\rangle)$

Nevertheless, the "continuous" predicates are not necessarily incompatible with the proposed LSTA model. Using LSTAs to automate the reasoning over quantum program against "continuous" predicates would be an interesting and fruitful research direction.

## ACKNOWLEDGEMENTS

We thank Mingsheng Ying for his insightful comments and suggestions on the potential of using automata to automate the reasoning based on "continuous" predicates. Moreover, we thank the anonymous reviewers for their feedback that helped to improve the quality of the paper. This work was supported by the Czech Ministry of Education, Youth and Sports ERC.CZ project LL1908, the Czech Science Foundation project 23-07565S, the FIT BUT internal project FIT-S-23-8151, National Science and Technology Council, R.O.C., projects NSTC-112-2222-E-001-002-MY3, NSTC-113-2119-M-001-009-, and NSTC-113-2222-E-027-007, Foxconn research project 05T-1120327-1C-, the Academia Sinica Grand Challenge Seeding Project, and The Swedish Research Council.

## REFERENCES

[1] Parosh Aziz Abdulla, Johanna Högberg, and Lisa Kaati. Bisimulation minimization of tree automata. *Int. J. Found. Comput. Sci.*, 18(4):699–713, 2007.

[2] Parosh Aziz Abdulla, Bengt Jonsson, Marcus Nilsson, and Mayank Saksena. A survey of regular model checking. In *International Conference on Concurrency Theory*, pages 35–48. Springer, 2004.

[3] Thorsten Altenkirch and Jonathan Grattage. A functional quantum programming language. In *20th IEEE Symposium on Logic in Computer Science (LICS 2005), 26-29 June 2005, Chicago, IL, USA, Proceedings*, pages 249–258. IEEE Computer Society, 2005.

[4] Matthew Amy. Towards large-scale functional verification of universal quantum circuits. In Peter Selinger and Giulio Chiribella, editors, *Proceedings 15th International Conference on Quantum Physics and Logic, QPL 2018, Halifax, Canada, 3-7th June 2018*, volume 287 of *EPTCS*, pages 1–21, 2018.

[5] MD SAJID ANIS, Abby-Mitchell, Héctor Abraham, et al. Qiskit: An open-source framework for quantum computing, 2021.

[6] Frank Arute et al. Quantum supremacy using a programmable superconducting processor. *Nature*, 574(7779):505–510, October 2019. Number: 7779 Publisher: Nature Publishing Group.

[7] Fabian Bauer-Marquart, Stefan Leue, and Christian Schilling. symqv: Automated symbolic verification of quantum programs. In Marsha Chechik, Joost-Pieter Katoen, and Martin Leucker, editors, *Formal Methods - 25th International Symposium, FM 2023, Lübeck, Germany, March 6-10, 2023, Proceedings*, volume 14000 of *Lecture Notes in Computer Science*, pages 181–198. Springer, 2023.

[8] Ethan Bernstein and Umesh V. Vazirani. Quantum complexity theory. In S. Rao Kosaraju, David S. Johnson, and Alok Aggarwal, editors, *Proceedings of the Twenty-Fifth Annual ACM Symposium on Theory of Computing, May 16-18, 1993, San Diego, CA, USA*, pages 11–20. ACM, 1993.

[9] Yves Bertot and Pierre Castéran. *Interactive theorem proving and program development: Coq'Art: the calculus of inductive constructions*. Springer Science & Business Media, 2013.

[10] Jacob D. Biamonte, Peter Wittek, Nicola Pancotti, Patrick Rebentrost, Nathan Wiebe, and Seth Lloyd. Quantum machine learning. *Nature*, 549(7671):195–202, 2017.

[11] Bruno Bogaert and Sophie Tison. Equality and disequality constraints on direct subterms in tree automata. In Alain Finkel and Matthias Jantzen, editors, *STACS 92, 9th Annual Symposium on Theoretical Aspects of Computer Science, Cachan, France, February 13-15, 1992, Proceedings*, volume 577 of *Lecture Notes in Computer Science*, pages 161–171. Springer, 1992.

[12] Ahmed Bouajjani, Peter Habermehl, Adam Rogalewicz, and Tomáš Vojnar. Abstract regular (tree) model checking. *International Journal on Software Tools for Technology Transfer*, 14(2):167–191, 2012.

[13] Lukas Burgholzer and Robert Wille. Advanced equivalence checking for quantum circuits. *IEEE Transactions on Computer-Aided Design of Integrated Circuits and Systems*, 40(9):1810–1824, 2020.

[14] Yudong Cao, Jonathan Romero, Jonathan P. Olson, Matthias Degroote, Peter D. Johnson, Mária Kieferová, Ian D. Kivlichan, Tim Menke, Borja Peropadre, Nicolas P. D. Sawaya, Sukin Sim, Libor Veis, and Alán Aspuru-Guzik. Quantum chemistry in the age of quantum computing. *Chemical Reviews*, 119(19):10856–10915, 2019. PMID: 31469277.

[15] Christophe Chareton, Sébastien Bardin, François Bobot, Valentin Perrelle, and Benoît Valiron. An automated deductive verification framework for circuit-building quantum programs. In Nobuko Yoshida, editor, *ESOP*, volume 12648 of *LNCS*, pages 148–177, Cham, March 2021. Springer International Publishing.

[16] Tian-Fu Chen, Jie-Hong R. Jiang, and Min-Hsiu Hsieh. Partial equivalence checking of quantum circuits. In *IEEE International Conference on Quantum Computing and Engineering, QCE 2022, Broomfield, CO, USA, September 18-23, 2022*, pages 594–604. IEEE, 2022.

[17] Yu-Fang Chen, Kai-Min Chung, Ondrej Lengál, Jyun-Ao Lin, and Wei-Lun Tsai. AutoQ: An automata-based quantum circuit verifier. In Constantin Enea and Akash Lal, editors, *Computer Aided Verification - 35th International Conference, CAV 2023, Paris, France, July 17-22, 2023, Proceedings, Part III*, volume 13966 of *Lecture Notes in Computer Science*, pages 139–153. Springer, 2023.

[18] Yu-Fang Chen, Kai-Min Chung, Ondrej Lengál, Jyun-Ao Lin, Wei-Lun Tsai, and Di-De Yen. An automata-based framework for verification and bug hunting in quantum circuits. *Proc. ACM Program. Lang.*, 7(PLDI):1218–1243, 2023.

[19] Yu-Fang Chen, Philipp Rümmer, and Wei-Lun Tsai. A theory of cartesian arrays (with applications in quantum circuit verification). In *International Conference on Automated Deduction*, pages 170–189. Springer, 2023.

[20] Carlo Ciliberto, Mark Herbster, Alessandro Davide Ialongo, Massimiliano Pontil, Andrea Rocchetto, Simone Severini, and Leonard Wossnig. Quantum machine learning: A classical perspective. *Proceedings of the Royal Society A: Mathematical, Physical and Engineering Sciences*, 474(2209), January 2018.

[21] Bob Coecke and Ross Duncan. Interacting quantum observables: categorical algebra and diagrammatics. *New Journal of Physics*, 13(4):043016, apr 2011.

[22] Hubert Comon, Max Dauchet, Rémi Gilleron, Florent Jacquemard, Denis Lugiez, Christof Löding, Sophie Tison, and Marc Tommasi. Tree automata techniques and applications, 2008.

[23] Ellie D'Hondt and Prakash Panangaden. Quantum weakest preconditions. *Mathematical Structures in Computer Science*, 16(3):429–451, 2006.

[24] Javier Esparza and Michael Blondin. *Automata Theory: An Algorithmic Approach*. MIT Press, 2023.

[25] Yuan Feng and Mingsheng Ying. Quantum Hoare logic with classical variables. *ACM Transactions on Quantum Computing*, 2(4):1–43, 2021.

[26] Emmanuel Filiot, Jean-Marc Talbot, and Sophie Tison. Tree automata with global constraints. *Int. J. Found. Comput. Sci.*, 21(4):571–596, 2010.

[27] Alexander S. Green, Peter LeFanu Lumsdaine, Neil J. Ross, Peter Selinger, and Benoît Valiron. Quipper: a scalable quantum programming language. In Hans-Juergen Boehm and Cormac Flanagan, editors, *ACM SIGPLAN Conference on Programming Language Design and Implementation, PLDI '13, Seattle, WA, USA, June 16-19, 2013*, pages 333–342. ACM, 2013.

[28] Daniel M. Greenberger, Michael A. Horne, and Anton Zeilinger. Going beyond Bell's theorem. In Menas Kafatos, editor, *Bell's Theorem, Quantum Theory and Conceptions of the Universe*, pages 69–72. Springer Netherlands, Dordrecht, 1989.

[29] Lov K. Grover. A fast quantum mechanical algorithm for database search. In Gary L. Miller, editor, *Proceedings of the Twenty-Eighth Annual ACM Symposium on the Theory of Computing, Philadelphia, Pennsylvania, USA, May 22-24, 1996*, pages 212–219. ACM, 1996.

[30] Stuart Hadfield. On the representation of Boolean and real functions as Hamiltonians for quantum computing. *ACM Transactions on Quantum Computing*, 2(4), dec 2021.

[31] Xin Hong, Xiangzhen Zhou, Sanjiang Li, Yuan Feng, and Mingsheng Ying. A tensor network based decision diagram for representation of quantum circuits. *ACM Trans. Design Autom. Electr. Syst.*, 27(6):60:1–60:30, 2022.

[32] Florent Jacquemard, Francis Klay, and Camille Vacher. Rigid tree automata. In Adrian-Horia Dediu, Armand-Mihai Ionescu, and Carlos Martín-Vide, editors, *Language and Automata Theory and Applications, Third International Conference, LATA 2009, Tarragona, Spain, April 2-8, 2009. Proceedings*, volume 5457 of *Lecture Notes in Computer Science*, pages 446–457. Springer, 2009.

[33] Florent Jacquemard, Francis Klay, and Camille Vacher. Rigid tree automata and applications. *Inf. Comput.*, 209(3):486–512, 2011.

[34] James C. King. Symbolic execution and program testing. *Commun. ACM*, 19(7):385–394, 1976.

[35] Florent Kirchner, Nikolai Kosmatov, Virgile Prevosto, Julien Signoles, and Boris Yakobowski. Frama-c: A software analysis perspective. *Formal aspects of computing*, 27:573–609, 2015.

[36] K Rustan M Leino. Dafny: An automatic program verifier for functional correctness. In *International conference on logic for programming artificial intelligence and reasoning*, pages 348–370. Springer, 2010.

[37] Ang Li and Sriram Krishnamoorthy. Sv-sim: Scalable pgas-based state vector simulation of quantum circuits. In *Proceedings of the International Conference for High Performance Computing, Networking, Storage and Analysis*, 2021.

[38] Junyi Liu, Bohua Zhan, Shuling Wang, Shenggang Ying, Tao Liu, Yangjia Li, Mingsheng Ying, and Naijun Zhan. Formal verification of quantum algorithms using quantum Hoare logic. In *International conference on computer aided verification*, pages 187–207. Springer, 2019.

[39] Igor L Markov and Yaoyun Shi. Simulating quantum computation by contracting tensor networks. *SIAM Journal on Computing*, 38(3):963–981, 2008.

[40] Sam McArdle, Suguru Endo, Alán Aspuru-Guzik, Simon C. Benjamin, and Xiao Yuan. Quantum computational chemistry. *Rev. Mod. Phys.*, 92:015003, Mar 2020.

[41] D. Michael Miller and Mitchell A. Thornton. QMDD: A decision diagram structure for reversible and quantum circuits. In *36th IEEE International Symposium on Multiple-Valued Logic (ISMVL 2006), 17-20 May 2006, Singapore*, page 30. IEEE Computer Society, 2006.

[42] Nikolaj Moll, Panagiotis Barkoutsos, Lev S Bishop, Jerry M Chow, Andrew Cross, Daniel J Egger, Stefan Filipp, Andreas Fuhrer, Jay M Gambetta, Marc Ganzhorn, Abhinav Kandala, Antonio Mezzacapo, Peter Müller, Walter Riess, Gian Salis, John Smolin, Ivano Tavernelli, and Kristan Temme. Quantum optimization using variational algorithms on near-term quantum devices. *Quantum Science and Technology*, 3(3):030503, jun 2018.

[43] P. Müller, M. Schwerhoff, and A. J. Summers. Automatic verification of iterated separating conjunctions using symbolic execution. In S. Chaudhuri and A. Farzan, editors, *Computer Aided Verification (CAV)*, volume 9779 of *LNCS*, pages 405–425. Springer-Verlag, 2016.

[44] Michael A. Nielsen and Isaac L. Chuang. *Quantum Computation and Quantum Information: 10th Anniversary Edition*. Cambridge University Press, USA, 10th edition, 2011.

[45] Tobias Nipkow, Lawrence C Paulson, and Markus Wenzel. *Isabelle/HOL: a proof assistant for higher-order logic*, volume 2283. Springer Science & Business Media, 2002.

[46] Simon Perdrix. Quantum entanglement analysis based on abstract interpretation. In *International Static Analysis Symposium*, pages 270–282. Springer, 2008.

[47] Helmut Seidl and Andreas Reuß. Extending $\mathcal{H}_1$-clauses with path disequalities. In Lars Birkedal, editor, *Foundations of Software Science and Computational Structures - 15th International Conference, FOSSACS 2012, Held as Part of the European Joint Conferences on Theory and Practice of Software, ETAPS 2012, Tallinn, Estonia, March 24 - April 1, 2012. Proceedings*, volume 7213 of *Lecture Notes in Computer Science*, pages 165–179. Springer, 2012.

[48] Meghana Sistla, Swarat Chaudhuri, and Thomas W. Reps. Symbolic quantum simulation with Quasimodo. In Constantin Enea and Akash Lal, editors, *Computer Aided Verification - 35th International Conference, CAV 2023, Paris, France, July 17-22, 2023, Proceedings, Part III*, volume 13966 of *Lecture Notes in Computer Science*, pages 213–225. Springer, 2023.

[49] Dimitrios Thanos, Tim Coopmans, and Alfons Laarman. Fast equivalence checking of quantum circuits of Clifford gates. In Étienne André and Jun Sun, editors, *Automated Technology for Verification and Analysis - 21st International Symposium, ATVA 2023, Singapore, October 24-27, 2023, Proceedings, Part II*, volume 14216 of *Lecture Notes in Computer Science*, pages 199–216. Springer, 2023.

[50] Yuan-Hung Tsai, Jie-Hong R. Jiang, and Chiao-Shan Jhang. Bit-slicing the Hilbert space: Scaling up accurate quantum circuit simulation. In *58th ACM/IEEE Design Automation Conference, DAC 2021, San Francisco, CA, USA, December 5-9, 2021*, pages 439–444. IEEE, 2021.

[51] Dominique Unruh. Quantum Hoare logic with ghost variables. In *2019 34th Annual ACM/IEEE Symposium on Logic in Computer Science (LICS)*, pages 1–13. IEEE, 2019.

[52] C. H. Valahu, V. C. Olaya-Agudelo, R. J. MacDonell, T. Navickas, A. D. Rao, M. J. Millican, J. B. Pérez-Sánchez, J. Yuen-Zhou, M. J. Biercuk, C. Hempel, T. R. Tan, and I. Kassal. Direct observation of geometric-phase interference in dynamics around a conical intersection. *Nature Chemistry*, 15:1503–1508, 2023.

[53] Guifré Vidal. Efficient classical simulation of slightly entangled quantum computations. *Phys. Rev. Lett.*, 91:147902, Oct 2003.

[54] Lieuwe Vinkhuijzen, Thomas Grurl, Stefan Hillmich, Sebastiaan Brand, Robert Wille, and Alfons Laarman. Efficient implementation of LIMDDs for quantum circuit simulation. In Georgiana Caltais and Christian Schilling, editors, *Model Checking Software - 29th International Symposium, SPIN 2023, Paris, France, April 26-27, 2023, Proceedings*, volume 13872 of *Lecture Notes in Computer Science*, pages 3–21. Springer, 2023.

[55] Chun-Yu Wei, Yuan-Hung Tsai, Chiao-Shan Jhang, and Jie-Hong R. Jiang. Accurate BDD-based unitary operator manipulation for scalable and robust quantum circuit verification. In Rob Oshana, editor, *DAC '22: 59th ACM/IEEE Design Automation Conference, San Francisco, California, USA, July 10 - 14, 2022*, pages 523–528. ACM, 2022.

[56] R. Wille, D. Große, L. Teuber, G. W. Dueck, and R. Drechsler. RevLib: An online resource for reversible functions and reversible circuits. In *Int'l Symp. on Multiple-Valued Logic*, pages 220–225, 2008. RevLib is available at http://www.revlib.org.

[57] Robert Wille, Rod Van Meter, and Yehuda Naveh. IBM's qiskit tool chain: Working with and developing for real quantum computers. In Jürgen Teich and Franco Fummi, editors, *Design, Automation & Test in Europe Conference & Exhibition, DATE 2019, Florence, Italy, March 25-29, 2019*, pages 1234–1240. IEEE, 2019.

[58] Mingkuan Xu, Zikun Li, Oded Padon, Sina Lin, Jessica Pointing, Auguste Hirth, Henry Ma, Jens Palsberg, Alex Aiken, Umut A Acar, et al. Quartz: superoptimization of quantum circuits. In *Proceedings of the 43rd ACM SIGPLAN International Conference on Programming Language Design and Implementation*, pages 625–640, 2022.

[59] Mingsheng Ying. Floyd-Hoare logic for quantum programs. *ACM Transactions on Programming Languages and Systems (TOPLAS)*, 33(6):1–49, 2012.

[60] Yordan S. Yordanov, David R. M. Arvidsson-Shukur, and Crispin H. W. Barnes. Efficient quantum circuits for quantum computational chemistry. *Phys. Rev. A*, 102:062612, Dec 2020.

[61] Nengkun Yu and Jens Palsberg. Quantum abstract interpretation. In *Proceedings of the 42nd ACM SIGPLAN International Conference on Programming Language Design and Implementation*, pages 542–558, 2021.

[62] Li Zhou, Nengkun Yu, and Mingsheng Ying. An applied quantum Hoare logic. In *Proceedings of the 40th ACM SIGPLAN Conference on Programming Language Design and Implementation*, pages 1149–1162, 2019.

[63] Alwin Zulehner and Robert Wille. Advanced simulation of quantum computations. *IEEE Trans. Comput. Aided Des. Integr. Circuits Syst.*, 38(5):848–859, 2019.

## A    SUPPLEMENTARY MATERIALS

### A.1    Proofs of Theorems in Section 5

We provide in this section the proofs of Theorems in Section 5.

THEOREM 8.   $\mathcal{L}(U_t(\mathcal{A})) = \{U_t(T) \mid T \in \mathcal{L}(\mathcal{A})\}$ *and* $|U_t(\mathcal{A})| = O(|\mathcal{A}|^2)$.

PROOF. Let $U = \begin{pmatrix} u_1 & u_2 \\ u_3 & u_4 \end{pmatrix}$. Since the algorithm is basically the product construction of LSTA's, the statement that $|U_t(\mathcal{A})| = |\mathcal{A}|^2$ in worst case is obvious. To show the correctness, we are going to show that there is a one-to-one correspondence between $\mathcal{L}(\mathcal{A})$ and $\mathcal{L}(\mathcal{A}')$. Since the transitions located at level $\leq t$ remain the same, we may, without loss of generality, assume that $t = 1$.

<u>Claim 1:</u> $U_1(\mathcal{A})$ *is an LSTA, i.e.,*

$$\forall \delta_1' \neq \delta_2' \in \Delta' \wedge \mathsf{top}(\delta_1') = \mathsf{top}(\delta_2') \implies \ell(\delta_1') \cap \ell(\delta_2') = \emptyset. \tag{4}$$

<u>Proof:</u> Since

(i) $\Delta' = \Delta_{=1}' \cup \Delta_{>1}' \cup \Delta_0'$
(ii) $\delta' \in \Delta_{=1}' \implies \mathsf{top}(\delta') \in Q$
(iii) $\delta' \in \Delta_{>1}' \cup \Delta_0' \implies \mathsf{top}(\delta') \in Q \times Q \times \{L, R\}$
(iv) $Q \cap (Q \times Q \times \{L, R\}) = \emptyset$

we only need to consider the following cases:

- Case $\delta_1', \delta_2' \in \Delta_{=1}'$: Since $\Delta'$ and $\Delta$ are in one-to-one correspondence, there are $\delta_1, \delta_2 \in \Delta_{=1}$ with $\delta_1 \neq \delta_2$ and $\mathsf{top}(\delta_1) = \mathsf{top}(\delta_2)$ such that $\ell(\delta_i') = \ell(\delta_i)$ for $i = 1, 2$. Thus

$$\ell(\delta_1') \cap \ell(\delta_2') = \ell(\delta_1) \cap \ell(\delta_2) = \emptyset.$$

- Case $\delta_1', \delta_2' \in \Delta_{>1}'$: By construction, $\delta_j'$ is of the form

$$(q^1, q^2, \frac{L}{R}) \,{-}C_j^1 \cap C_j^2{\to}\, x^j((q_j^{1l}, q_j^{2l}, \frac{L}{R}), (q_j^{1r}, q_j^{2r}, \frac{L}{R}))$$

with the corresponding transitions

$$\delta_j^1 = q^1 \,{-}C_j^1{\to}\, x^j((q_j^{1l}, q_j^{1r})), \quad \delta_j^2 = q^2 \,{-}C_j^2{\to}\, x^j((q_j^{2l}, q_j^{2r}))$$

in $\Delta$, $j = 1, 2$. Suppose that $\ell(\delta_1') \cap \ell(\delta_2') \neq \emptyset$. Then

$$C_1^1 \cap C_1^2 \cap C_2^1 \cap C_2^2 \neq \emptyset.$$

Since $\mathcal{A}$ is an LSTA and $\mathsf{top}(\delta_1^1) = \mathsf{top}(\delta_2^1)$, $\mathsf{top}(\delta_1^2) = \mathsf{top}(\delta_2^2)$, we have both $\delta_1^1 = \delta_2^1$ and $\delta_1^2 = \delta_2^2$. Thus $\delta_1' = \delta_2'$ and it leads to a contradiction. Hence we have $\ell(\delta_1') \cap \ell(\delta_2') = \emptyset$.
- In general, the case $\delta_1', \delta_2' \in \Delta_{>t}' \cup \Delta_0'$ follows by the same argument as above.

Thus, we have $U_1(\mathcal{A})$ is an LSTA.    ∎

We are going to construct a bijective map

$$O_{U_1} : \mathcal{L}(\mathcal{A}) \to \mathcal{L}(\mathcal{A}')$$

$$T_\rho \mapsto T_{\rho'}'.$$

The construction is rather straightforward: let $T := T_\rho \in \mathcal{L}(\mathcal{A})$ with an associated accepting run $\rho$. For $v \in \{0, 1\}^j$, $j = 0, 1, \ldots, n - 1$, we denote by $q = q_\epsilon := \mathsf{top}(\rho(\epsilon)) \in \mathcal{R}$ its root state and $q_v := \mathsf{top}(\rho(v)) \in Q$ for each node $v$. Construct a tree $T' := T_{\rho'}'$ with an associated run $\rho'$ inductively as follows:

(i) $\text{dom}(T') = \text{dom}(T)$

(ii) <u>Base</u>: Since $\mathcal{R}' = \mathcal{R}$, we set $q'_\epsilon = q$. Since there is a one-to-one correspondence between $\Delta_{=1}$ and $\Delta'_{=1}$ and the transition in $\Delta'_{=1}$ corresponding to $\rho(\epsilon)$ is

$$q \ -C_{\rho,\epsilon} \rightarrow x_1((q_0,q_1,L),(q_0,q_1,R)).$$

It is well-defined to set $\rho'(\epsilon)$ to be the above transition.

(iii) <u>Inductive</u>: For any other internal node $v \in \{0,1\}^j$, $j = 1,\ldots,n-1$, the transition

$$\rho'(v) = q'_v \ -C_{\rho',v} \rightarrow T'(v)(q'_{v.0}, q'_{v.1})$$

can be defined inductively on $j$ as follows: suppose that $\rho'(w)$, $w \in \{0,1\}^j$, $j = 0,1,\ldots k-1$ have been defined. Then $\text{top}(\rho'(v))$ has been determined for all $v \in \{0,1\}^k$ and $\text{top}(\rho'(v)) = q'_v$ is of the form $(q_{v_1}, q_{v_2}, \frac{L}{R})$ for some $v_1, v_2 \in \{0,1\}^k$. We then set $\rho'(v)$ to be the transition in $\Delta'$ corresponding to $\rho(v_1)$ and, $\rho(v_2)$ which also determines $\text{top}(\rho'(v.0))$ and $\text{top}(\rho'(v.1))$.

(iv) Finally, for any leaf node $u \in \{0,1\}^n$, since $\text{top}(\rho'(u))$ has been defined, say

$$q'_u = \text{top}(\rho'(u)) = (q_{u^1}, q_{u^2}, \frac{L}{R})$$

for some $u^1, u^2 \in \{0,1\}^n$, and there are corresponding leaf transitions $\rho(u^1) = q_{u^1} \ -C_{\rho,u^1} \rightarrow a$ and $\rho(u^2) = q_{u^2} \ -C_{\rho,u^2} \rightarrow b$, we set the leaf transition

$$\rho'(u) := \begin{cases} q'_u \ -C_{\rho,u^1} \cap C_{\rho,u^2} \rightarrow u_1 a + u_2 b & \text{if } q'_u = (q_{u^1}, q_{u^2}, L) \\ q'_u \ -C_{\rho,u^1} \cap C_{\rho,u^2} \rightarrow u_3 a + u_4 b & \text{if } q'_u = (q_{u^1}, q_{u^2}, R) \end{cases}$$

in $\Delta'_0$.

<u>Injectivity</u>: Obviously the tree $T' = T'_{\rho'}$ with $\rho'$ being a run of the LSTA $\langle Q', \Sigma', \Delta', \mathcal{R} \rangle$ and the map is one-to-one by the construction. The functionality $T' = U_1(T)$ can also be easily seen from the leaves. We are left to show that $\rho'$ is accepting, i.e.,

<u>Claim 2</u>: *For any $v \in \text{dom}(T')$, we have*

$$\bigcap_{\delta' \in \text{level}(\rho', \text{ht}(v))} \ell(\delta') \neq \emptyset \tag{5}$$

*or in other words, for any $i = 0, 1, \ldots, n$, we have*

$$\bigcap_{v \in \{0,1\}^i} \ell(\rho'(v)) \neq \emptyset \tag{6}$$

<u>Proof:</u> For $i = 0$ there is only one transition $\rho'(\epsilon)$. By construction, we have $\ell(\rho'(\epsilon)) = \ell(\rho(\epsilon)) \neq \emptyset$ where the later inequality holds since $\rho$ is an accepting run. For $i \in [n-1]$, since each $\rho'(v)$, $v \in \{0,1\}^i$, has option $\ell(\rho'(v)) = \ell(\rho(v_1)) \cap \ell(\rho(v_2))$ for some $v_1, v_2 \in \{0,1\}^i$ by construction, we have

$$\bigcap_{v \in \{0,1\}^i} \ell(\rho'(v)) \supseteq \bigcap_{v \in \{0,1\}^i} \ell(\rho(v)) \neq \emptyset,$$

where again the later inequality holds since $\rho$ is an accepting run of $\mathcal{A}$. The same argument also holds for the case of leaf level $i = n$.

∎

Hence we conclude that $\rho'$ is an accepting run by the LSTA $\mathcal{A}'$.

<u>Surjectivity.</u> To show the surjectivity of $O_{U_1}$, let $T' := T'_{\rho'} \in \mathcal{L}(\mathcal{A}')$ and $\rho'$ be the accepting run of $\mathcal{A}'$ over $T'$. We are going to construct a tree $T := T_\rho$ with its associated run $\rho$ accepted by $\mathcal{A}$ such that $O_{U_1}(T_\rho) = T'_{\rho'}$. Similarly, let us simplify some notations: for internal nodes $v \in \{0,1\}^j$, $j = 0, 1, \ldots, n-1$ and leaf nodes $u \in \{0,1\}^n$, we denote by $q'_v := \text{top}(\rho'(v)) \in Q'$ for each node $v$ and $q := q'_\epsilon = \text{top}(\rho'(\epsilon))$. Let us construct a tree $T := T_\rho$ with run $\rho$ inductively as follows:

(i) <u>Base</u>: Since $\mathcal{R} = \mathcal{R}'$, we set $q_\epsilon := \text{top}(\rho'(\epsilon)) = q$. For $T(\epsilon) = T'(\epsilon) = x_1$, the transition $\rho'(\epsilon)$ is of the form

$$q -C_{\rho',\epsilon} \rightarrow x_1((q_l, q_r, L), (q_l, q_r, R))$$

for some $q_l, q_r \in Q$. By algorithm, there is a one-to-one correspondence between $\Delta_{=1}$ and $\Delta'_{=1}$ and the transition corresponding to $\delta'_\epsilon$ is $q -C_{\rho',\epsilon} \rightarrow x_1(q_l, q_r) \in \Delta$. We may set the transition $\rho(\epsilon)$ to be the above one, which induces $q_0 = q_l$ and $q_1 = q_r$.

(ii) <u>Inductive</u>: For any other internal node $v \in \{0,1\}^j$, $j = 1, \ldots, n-1$, the transitions

$$\rho(v) := q_v -C_{\rho,v} \rightarrow T(v)(q_{v.0}, q_{v.1})$$

can be constructed inductively on $j$ as follows. Suppose that $\rho(w), w \in \{0,1\}^j, j = 0, 1, \ldots, k-1$ have been defined for some $k < n$. Let $v \in \{0,1\}^k$. Then $q_v = \text{top}(\rho(v))$ has been determined too. From the algorithm there exists a $\hat{v} \in \{0,1\}^k$ with the corresponding transition $\rho'(\hat{v})$ being of the form

$$(q^1, q^2, \frac{L}{R}) -C_{\rho',\hat{v}} \rightarrow x_{k+1}((q_l^1, q_l^2, \frac{L}{R}), (q_r^1, q_r^2, \frac{L}{R}))$$

for some $q^1, q_l^1, q_r^1, q^2, q_l^2, q_r^2 \in Q$ such that $q_v = q^1 \vee q^2$ and there exist transitions

$$q^s -C_s \rightarrow T(v)(q_l^s, q_r^s) \in \Delta, \ \ s = 1, 2$$

such that $C_{\rho',v} = C_1 \cap C_2$. We may set

$$\delta_v := \begin{cases} q^1 -C_1 \rightarrow x_{k+1}(q_l^1, q_r^1) & \text{if } q_v = q^1 \\ q^2 -C_2 \rightarrow x_{k+1}(q_l^2, q^r) & \text{if } q_v = q^2 \end{cases}.$$

Note that it also determines $q_{v.0}$ and $q_{v.1}$.

(iii) Similarly, for any leaf node $u \in \{0,1\}^n$, the top $q_u$ of the desired transition $\rho(u)$ has been determined as above and by construction there exists an $\hat{u} \in \{0,1\}^n$ such that the corresponding leaf transition $\rho'(u)$ is of the form

$$(p^1, p^2, \frac{L}{R}) -C_{\rho',u} \rightarrow T'(u) = u_1 a + u_2 b$$

and $q_u = p^1 \vee p^2$. By algorithm, there are leaf transitions $p^1 -C_1 \rightarrow a$ and $p^2 -C_2 \rightarrow b$ in $\Delta_0$ such that $C_{\rho',u} = C_1 \cap C_2$. Thus, we may set

$$\delta_u = \begin{cases} p^1 -C_1 \rightarrow a & \text{if } q_u = p^1 \\ p^2 -C_2 \rightarrow b & \text{if } q_u = p^2 \end{cases}.$$

The run $\rho$ constructed above is obviously well-defined. We claim that $\rho$ is accepting by $\mathcal{A}$, i.e., <u>Claim 3</u>: *For each $i = 0, 1, \ldots, n$, we have*

$$\bigcap_{v \in \{0,1\}^i} \ell(\rho(v)) \neq \emptyset. \tag{7}$$

<u>Proof</u>: For $i = 0$, there is only one transition $\rho(\epsilon)$. By construction we have $\ell(\rho(\epsilon)) = \ell(\rho'(\epsilon)) \neq \emptyset$ where the later inequality holds since $\rho'$ is accepting. For $i \in [n-1]$, by construction, each $\rho(v)$ has option $\ell(\rho(v)) \supseteq \ell(\rho'(v'))$ for some $v' \in \{0,1\}^i$. Thus

$$\bigcap_{v \in \{0,1\}^i} \ell(\rho(v)) \supseteq \bigcap_{v' \in \{0,1\}^i} \ell(\rho'(v')) \neq \emptyset$$

where again the later inequality holds since $\rho'$ is accepting. The same argument also works for leaf level $i = n$.    ∎

□

<u>Claim 4</u>: *Every accepting run $\rho$ for $\mathcal{A}$, there exists an accepting run $\rho'$ for $U_t(\mathcal{A})$, such that*

- $sym \circ \rho = T \wedge sym \circ \rho' = U_t(T)$
- $top(\rho(\epsilon)) = top(\rho'(\epsilon))$
- $\bigcap_{\delta \in \text{level}(\rho, d)} \ell(\delta) = \bigcap_{\delta \in \text{level}(\rho', d)} \ell(\delta)$ *for all $d$*

<u>Proof:</u>

∎

THEOREM 9. $\mathcal{L}(\text{CU}_t^c(\mathcal{A})) = \{\text{CU}_t^c(T) \mid T \in \mathcal{L}(\mathcal{A})\}$ *and* $|\text{CU}_t^c(\mathcal{A})| = |\mathcal{A}| + |U_t(\mathcal{A})|$.

PROOF. By algorithm, the statement that $|\text{CU}_t^c(\mathcal{A})| = |\mathcal{A}| + |U_r(\mathcal{A})|$ in worst cases is obvious.
<u>Claim 5</u>: $\text{CU}_t^c(\mathcal{A})$ *is an LSTA.*

<u>Proof:</u> Since $Q' \cap Q^U = \emptyset$, we have $\text{top}(\delta') \neq \text{top}(\delta^U)$ for any pair $\delta' \in \Delta'$ and $\delta^U \in \Delta^U$. Thus we only need to consider the following cases

- $\forall \delta_1' \neq \delta_2' \in \Delta' : \text{top}(\delta_1') = \text{top}(\delta_2') \implies \ell(\delta_1') \cap \ell(\delta_2') = \emptyset$
- $\forall \delta_1^U \neq \delta_2^U \in \Delta^U : \text{top}(\delta_1^U) = \text{top}(\delta_2^U) \implies \ell(\delta_1^U) \cap \ell(\delta_2^U) = \emptyset$.

However the former follows by the fact that $\mathcal{A}$ is LSTA and the later follows from the proof of Theorem 8, i.e., $U_t(\mathcal{A})$ is LSTA.    ∎

By the proof of Theorem 8, there is a one-to-one correspondence between the accepting runs $\rho$ of $\mathcal{A}$ and runs $\rho^U$ of $U_t(\mathcal{A})$. The construction of the one-to-one correspondence between the runs $\rho$ of $\mathcal{A}$ and the runs $\rho^{CU}$ of $\text{CU}_t^c(\mathcal{A})$ can be done in an obvious way. By algorithm, we have, for each $i = 0, 1, \ldots, n$,

$$\bigcap_{v \in \{0,1\}^i} \ell(\rho^{CU}(v)) \supseteq \bigcap_{v \in \{0,1\}^i} \ell(\rho^U(v)) = \bigcap_{v \in \{0,1\}^i} \ell(\rho(v)) \neq \emptyset.$$

That is, $\rho^{CU}$ is accepting. Hence the Theorem follows.    □

THEOREM 10. $\mathcal{L}(\text{X}_t(\mathcal{A})) = \{\text{X}_t(T) \mid T \in \mathcal{L}(\mathcal{A})\}$ *and* $|\text{X}_t(\mathcal{A})| = |\mathcal{A}|$.

PROOF. From the algorithm, it is obvious to see that $\text{X}_t(\mathcal{A})$ is LSTA and $|\text{X}_t(\mathcal{A})| = |\mathcal{A}|$. The construction of one-to-one correspondence between runs $\rho$ of $\mathcal{A}$ and the runs $\rho'$ of $\text{X}_t(\mathcal{A})$ is essentially the same as in the proof of [18]. Moreover, we have $\ell(\rho(v)) = \ell(\rho'(v))$ for each node $v$, hence $\rho'$ is accepting if and only if $\rho$ is accepting.    □

THEOREM 11. $\mathcal{L}(\text{D}_t^{r_0, r_1}(\mathcal{A})) = \{\text{D}_t^{r_0, r_1}(T) \mid T \in \mathcal{L}(\mathcal{A})\}$ *and* $|\text{D}_t^{r_0, r_1}(\mathcal{A})| = 2|\mathcal{A}|$.

PROOF. Since $Q \cap Q' = \emptyset$, by the same reason as in the proof of Theorem 9, $\text{D}_t^{r_0, r_1}(\mathcal{A})$ is indeed an LSTA. The correctness follows from the same argument as in the proof of Theorem 10.    □

## A.2    Proofs for Theorems in Section 8

THEOREM 18. $\mathcal{L}(\text{CX}(n)(\mathcal{A})) = \{\text{CX}_n^{n-1}(\cdots \text{CX}_2^1(T) \cdots) \mid T \in \mathcal{L}(\mathcal{A})\}$ *and* $|\text{CX}(n)(\mathcal{A})| = 2|\mathcal{A}|$.

THEOREM 19. $\mathcal{L}(\text{CX}(n)^{-1}(\mathcal{A})) = \{\text{CX}_2^1(\cdots \text{CX}_n^{n-1}(T) \cdots) \mid T \in \mathcal{L}(\mathcal{A})\}$ *and* $|\text{CX}(n)^{-1}(\mathcal{A})| = 2|\mathcal{A}|$.

We will prove the above Theorems together in the following.

PROOF. It is straightforward to see the complexities of both algorithms.

Since $\overline{\mathcal{A}}$ is obtained by swapping left and right children of all the transitions in $\mathcal{A}$, there is a one-to-one correspondence between the accepting runs $\rho$ in $\mathcal{L}(\mathcal{A})$ and runs $\overline{\rho}$ in $\mathcal{L}(\overline{\mathcal{A}})$. For each run $\rho$ in $\mathcal{L}(\mathcal{A})$, the construction of the associated run $\rho'$ (resp. $\rho'_{-1}$) in $\mathcal{L}(\text{CX}(n)(\mathcal{A}))$ (resp. $\mathcal{L}(\text{CX}(n)^{-1}(\mathcal{A}))$) can be derived directly from Algorithm 6 (resp. from Algorithm 7) (e.g. in Fig. 23, where the state <span style="color:purple">$q$</span> in <span style="color:blue">blue</span> denotes the corresponding barred version $\overline{q}$).

The correctness of both algorithms follow from the observation: the effect of applying each $\text{CX}^i_{i+1}$ gate on a tree will swap the right-hand side subtrees just below the $i$-th level and keep the left-hand side subtrees unchanged. Moreover,

(i) the trees in $\mathcal{L}(\overline{\mathcal{A}})$ are obtained from flipping the trees in $\mathcal{L}(\mathcal{A})$ around the central vertical line and vice versa.

(ii) CX($n$): performing swaps from top to bottom can be seen as flipping subtrees around their central vertical lines.

(iii) CX($n$)$^{-1}$: performing swaps from bottom to top is equivalent to flipping subtrees around their central vertical lines level-by-level from top to bottom.

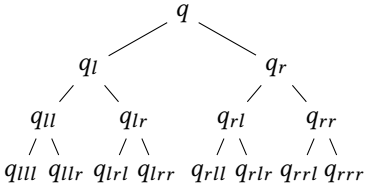

(a) A run in $\mathcal{L}(\mathcal{A})$

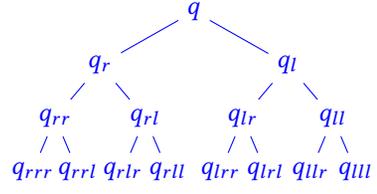

(b) The corresponding run in $\mathcal{L}(\overline{\mathcal{A}})$

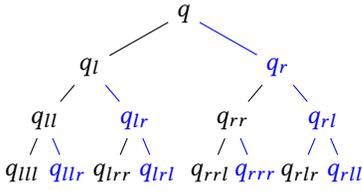

(c) The corresponding run in $\mathcal{L}(\text{CX}(n)(\mathcal{A}))$

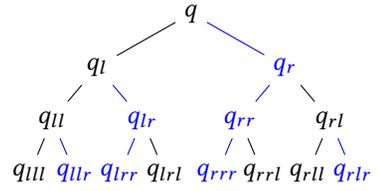

(d) The corresponding run in $\mathcal{L}(\text{CX}(n)^{-1}(\mathcal{A}))$

Fig. 23. One-to-one correspondence between runs in $\mathcal{L}(\mathcal{A})$, $\mathcal{L}(\overline{\mathcal{A}})$, CX($n$)($\mathcal{A}$) and CX($n$)$^{-1}(\mathcal{A})$

$\square$

### A.3 Verification of Diagonal Hamiltonian Simulation Circuit

In many areas, like *quantum chemistry* or *material science*, it is often important to figure out how a quantum system will evolve from some given initial state, for instance, a basis state [14, 40, 42, 52]. In these areas, the modeled systems can often be described using the so-called *Hamiltonians*, i.e., $2^n \times 2^n$ *Hermitian* matrices (a matrix is Hermitian if it is equal to its conjugate transpose) [40]. Any Hamiltonian can be decomposed into a linear combination of tensor products of identity and Pauli gates (I, X, Y, Z) [44, Chapter 4.7]. Hamiltonians are, in general, not unitary matrices, so they cannot be directly used as a part of a quantum circuit. The evolution of a system described by a Hamiltonian $\hat{H}$ in time $t$ can, however, be computed by a quantum circuit implementing the unitary operator $U(t) = e^{-it\hat{H}}$; this approach is called *Hamiltonian simulation* [44].

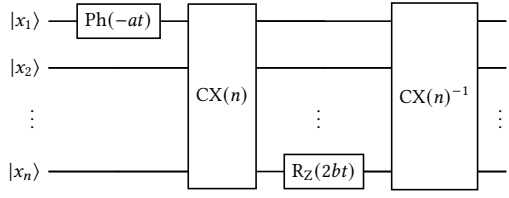

Fig. 24. A circuit implementing the unitary operator $U(t) = e^{-it(aI^{\otimes n} + bZ^{\otimes n})}$. In the experimental evaluation, we use $a = b = -\frac{\pi}{2}$ and $t = 1$.

$$
\begin{array}{llll}
q \; -C_r\!\to x(0_r, r_1) & q \; -C_l\!\to x(l, 0_l) & r_1 \; -\hat{C}_r\!\to x(\hat{0}_r, \hat{r}_2) & \hat{r}_2 \; -C\!\to 1 \\
r_1 \; -C_r\!\to x(0_r, r_2) & r_2 \; -C_l\!\to x(l, 0_l) & r_2 \; -\hat{C}_l\!\to x(\hat{l}, \hat{0}_l) & \hat{l} \; -C\!\to 1 \\
r_2 \; -C_r\!\to x(0_r, r_1) & 0_r \; -C_l\!\to x(0_l, 0_l) & l \; -\hat{C}_l\!\to x(\hat{l}, \hat{0}_l) & \hat{0}_l \; -C\!\to 0 \\
0_r \; -C_r\!\to x(0_r, 0_r) & l \; -C_l\!\to x(l, 0_l) & 0_r \; -\hat{C}_r\!\to x(\hat{0}_r, \hat{0}_r) & \hat{0}_r \; -C\!\to 0 \\
l \; -C_r\!\to x(0_r, r_1) & 0_l \; -C_l\!\to x(0_l, 0_l) & 0_l \; -\hat{C}_l\!\to x(\hat{0}_l, \hat{0}_l) & \\
0_l \; -C_r\!\to x(0_r, 0_r) & & &
\end{array}
$$

Fig. 25. The LSTA $\mathscr{A}_{even}$ with $\mathcal{L}(\mathscr{A}_{even}) = \{|x\rangle \mid n \in \mathbb{N} \wedge x \in \{0,1\}^n \wedge f_n(x) = 0\}$.

Our framework allows to express *parameterized Hamiltonian simulation* of certain *diagonal* Hamiltonians. A Hamiltonian $\hat{H}$ is diagonal if $\hat{H}$ is a diagonal matrix. The class of Hamiltonians that we support are those that can be described as a linear combination of the $n$-fold identity and Pauli Z gates, i.e., $\hat{H} = aI^{\otimes n} + bZ^{\otimes n}$ for $a, b \in \mathbb{R}$ and $n \in \mathbb{N}$ being the parameter. Such a Hamiltonian can be simulated by the unitary operator $U(t) = e^{-it\hat{H}}$ represented by the circuit in Fig. 24. Below, we give a formal derivation of correctness of the circuit and the results are given in Table 3.

Table 3. Results for verification of diagonal Hamiltonian simulation

| | | AutoQ | | |
|---|---|---|---|---|
| | #G | post$_C$ | $\subseteq$ | total |
| Fig. 24 | 4 | 0.0s | 0.0s | **0.0s** |

Consider the Hamiltonian $H_{f_n} = \frac{1}{2}(I^{\otimes n} - Z^{\otimes n})$, where $I^{\otimes n}$ is the $n$-fold tensor product of the identity matrix $\begin{pmatrix} 1 & 0 \\ 0 & 1 \end{pmatrix}$ and $Z^{\otimes n}$ is the $n$-fold tensor product of the Pauli operator $Z = \begin{pmatrix} 1 & 0 \\ 0 & -1 \end{pmatrix}$. $H_{f_n}$ actually represents the Boolean function $f_n \colon \{0,1\}^n \to \{0,1\}$ defined by $f_n(x_1, \ldots, x_n) = \bigoplus_{i=1}^n x_i$ in the sense that $H_{f_n} |x\rangle = f_n(x) |x\rangle$ for all $x \in \{0,1\}^n$ [30]. It is well-known that such a Hamiltonian $H_{f_n}$ can be simulated efficiently by a quantum circuit that implements the unitary operator $U(\gamma) = e^{-i\gamma H_{f_n}}$ and the standard implementation is showed in Fig. 24. Moreover, one can compute by hand that at time $\gamma = \pi$, we have

$$U(\pi) |x\rangle = (-1)^{f_n(x)} |x\rangle \tag{8}$$

for all $x \in \{0,1\}^n$. We demonstrate here how to verify Eq. (8) for all $n \in \mathbb{N}$ automatically via parameterized verification techniques. We start with the LSTA $\mathscr{A}_{even} = \langle Q, \Sigma, \Delta, \mathcal{R} = \{q\} \rangle$ in Fig. 25 representing the set of basis states $|x\rangle$ with $x \in \{0,1\}^n$ having even numbers of 1's. Then we execute the circuit Fig. 24 for $a = \frac{1}{2}, b = -\frac{1}{2}$ and $t = \pi$ as follows:

(i) we first unfold $\mathscr{A}_{even}$ from top-down and apply the global phase gate $\text{Ph}(-\frac{\pi}{2})$ on the first qubit;

(ii) fold the resulting LSTA and apply the CX($n$) operation via Algorithm 6;
(iii) unfold again from bottom-up and apply $R_Z(-\pi)$ on the last qubit;
(iv) fold the resulting LSTA and apply the CX($n$)$^{-1}$ operation via Algorithm 7.

After executing, we can then verify the property $U(\pi)(\mathcal{A}_{even}) \subseteq \mathcal{A}_{even}$ corresponding to Eq. (8) via the inclusion algorithm in Section 6 of LSTA's.

## A.4  More Algorithms for Parameterized Verification

Apart from the staircases CX operators in both directions, we have also implemented several other parameterized gates interpreting the quantum circuit operations in Fig. 26.

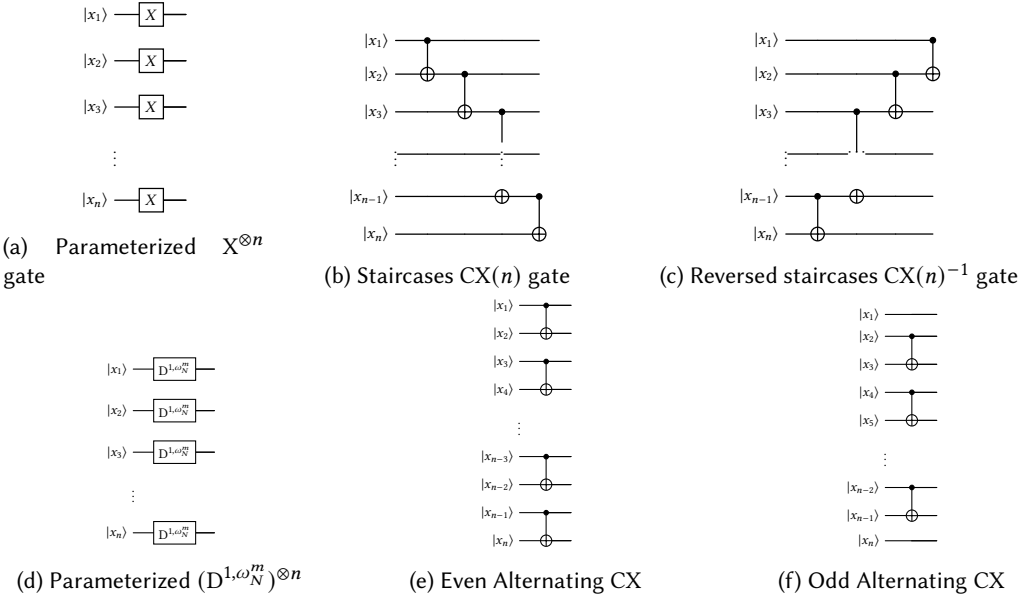

(a)  Parameterized  X$^{\otimes n}$ gate

(b)  Staircases CX($n$) gate

(c)  Reversed staircases CX($n$)$^{-1}$ gate

(d)  Parameterized ($D^{1,\omega_N^m}$)$^{\otimes n}$

(e)  Even Alternating CX

(f)  Odd Alternating CX

Fig. 26.  Parameterized gate operations

The Algorithm 8 corresponding to Fig. 26a is nothing but the $\overline{\mathcal{A}}$ used in Algorithms 6 and 7 and the correctness is obvious.

---

**Algorithm 8:** Algorithm for applying X on all qubits simultaneously as Fig. 26a

---

**Input:** A LSTA $\mathcal{A} = \langle Q, \Sigma, \Delta, \mathcal{R} \rangle$ representing parameterized set of quantum states and X
**Output:** A LSTA X$^{\otimes n}(\mathcal{A})$
1  $\Delta^R := \{\delta^R = q \xrightarrow{-C} f(q_r, q_l) \mid \delta = q \xrightarrow{-C} f(q_l, q_r) \in \Delta_{\neq 0}\} \cup \Delta_0$;
2  **return** $\mathcal{A}^R = \langle Q, \Sigma, \Delta^R, \mathcal{R} \rangle$;

---

The idea behind Algorithm 9 for alternating CX operations as Figs. 26e and 26f is simple: we use the same strategy as Algorithm 6. And the parity will be determined by choosing root states being in either the first or second copy of $\mathcal{A}$. The correctness follows similarly from the proof of Theorem 18.

We can also apply a fixed phase-shift gate $D^{1,\omega_N^m}$ simultaneously on all qubits as in Fig. 26d. The intuition behind Algorithm 10 is as follows:

---

**Algorithm 9:** Algorithm for applying CX gates alternating on odd Fig. 26f or even Fig. 26e qubits

---

**Input:** A LSTA $\mathcal{A} = \langle Q, \Sigma, \Delta, \mathcal{R} \rangle$ representing a parameterized quantum states. $b \in \{0, 1\}$ stands for even or odd parameterized CX.

**Output:** A LSTA $\text{CX}^{alt,b}(n)(\mathcal{A}) =: \mathcal{A}^R = \langle Q^R, \sigma, \Delta^R, \mathcal{R}^R \rangle$

1  $Q^R := \{q^i \mid q \in Q, i = 0, 1, 2\};$

2  $\Delta_{\neq 0}^R := \{q^0 - C \to f(q_l^1, q_r^2),\ q^1 - C \to f(q_l^0, q_r^0),\ q^2 - C \to f(q_r^0, q_l^0) \mid q - C \to f(q_l, q_r) \in \Delta_{\neq 0}\}$

   $\Delta_0^R := \{q^i - C \to k \mid q - C \to k \in \Delta_0, i = 0, 1, 2\};$

3  $\Delta^R := \Delta_{\neq 0}^R \cup \Delta_0^R;\ \mathcal{R}^R := \{q^b \mid q \in \mathcal{R}\};$

4  **return** $\mathcal{A}^R$

---

(i) we first make $N$ copies of $\mathcal{A}$ and for each $i$-th copy we modify the leaf symbols by multiplying $\omega_N^i$;

(ii) the right-hand side subtrees jump level by level from $i$-th copy to $i + 1$-th one.

The correctness can be derived in a similar way as in the proof of Theorem 18.

---

**Algorithm 10:** Algorithm for applying phase gate $\text{D}^{1,\omega_N}$ on all qubits simultaneously, Fig. 26d.

---

**Input:** A LSTA $\mathcal{A} = \langle Q, \Sigma, \Delta, \mathcal{R} \rangle$, $N \in \mathbb{N}$ and an $\text{D}^{1,\omega_N} = \begin{pmatrix} 1 & 0 \\ 0 & \omega_N \end{pmatrix}$.

**Output:** A LSTA $\text{D}^{1,\omega_N}(n)(\mathcal{A}) =: \mathcal{A}^R = \langle Q^R, \Sigma, \Delta^R, \mathcal{R} \rangle$

1  $Q^{\omega_N^i} := \{q^{\omega_N^i} \mid q \in Q\},\ Q^R := \bigcup_{i \in \mathbb{Z}/N\mathbb{Z}} Q^{\omega_N^i},\ \omega_N : Q^R \to Q^R$ by $\omega_N(q^{\omega_N^i}) = q^{\omega_N^{i+1}};$

2  For $i \in \mathbb{Z}/N\mathbb{Z}$,

   $$\Delta^{\omega_N^i} := \{\delta^{\omega_N^i} = q^{\omega_N^i} - C \to f((q_l)^{\omega_N^i}, (q_r)^{\omega_N^i}) \mid \delta = q - C \to f(q_l, q_r) \in \Delta_{\neq 0}\} \cup \Delta_0^{\omega_N^i}$$

   $$\Delta_0^{\omega_N^i} := \{\delta^{\omega_N^i} = q^{\omega_N^i} - C \to a \times \omega_N^i \mid \delta = q - C \to a \in \Delta_0\}.$$

   $$\Delta^R := \bigcup_{i \in \mathbb{Z}/N\mathbb{Z}} \{\delta = q - C \to f(q_l, (q_r)^{\omega_N}) \mid \delta_1 = q - C \to f(q_l, q_r) \in \Delta^{\omega_N^i}\} \cup \bigcup_{i \in \mathbb{Z}/N\mathbb{Z}} \Delta_0^{\omega_N^i}.$$

   **return** $\mathcal{A}^R$

---